\documentclass[article, prx, nobibnotes, secnumarabic, superscriptaddress, twocolumn, aps,longbibliography, nofootinbib]{revtex4-2}

\usepackage{graphicx} 
\usepackage{dcolumn} 
\usepackage{siunitx}
\usepackage[T1]{fontenc} 
\usepackage{xr-hyper}
\usepackage{hyperref}
\usepackage{amsmath}
\usepackage{amssymb}
\usepackage{amsfonts}
\usepackage{amsthm}
\usepackage{bm}

\hypersetup{colorlinks,breaklinks,
            urlcolor=[rgb]{0.54,0.08,0.08},
            linkcolor=[rgb]{0.544,0.08,0.08},
            citecolor=[rgb]{0.54,0.08,0.08},
            filecolor=[rgb]{0.54,0.08,0.08}}
\usepackage{soul}

\newcommand\blfootnote[1]{%
  \begingroup
  \renewcommand\thefootnote{}\footnote{#1}%
  \addtocounter{footnote}{-1}%
  \endgroup
}

\newcommand{\StanfordMSE}{Department of Materials Science and Engineering, Stanford University, Stanford, CA 94305, USA}
\newcommand{\StanfordP}{Department of Physics, Stanford University, Stanford, CA 94305, USA}
\newcommand{\StanfordAP}{Department of Applied Physics, Stanford University, Stanford, CA 94305, USA}
\newcommand{\SIMES}{Stanford Institute for Materials and Energy Sciences, SLAC National Accelerator Laboratory, Menlo Park, CA 94025, USA}

\newcommand{\GLAM}{Geballe Laboratory for Advanced Materials, Stanford University, Stanford, CA 94305, USA}

\begin{document}

\title{Bend Contour Electron Tomography (BCET): Quantitative strain and topography mapping}

\author{Henry~G.~Bell}
\thanks{These authors contributed equally: H.B. and C.X.}
\affiliation{\StanfordAP}
\affiliation{\SIMES}

\author{Chenhang~Xu}
\thanks{These authors contributed equally: H.B. and C.X.}
\affiliation{\StanfordAP}
\affiliation{\SIMES}

\author{Arthur~McCray}
\affiliation{\StanfordMSE}

\author{Minyong~Han}
\affiliation{\StanfordAP}
\affiliation{\SIMES}

\author{Yicheng~Zhuang}
\affiliation{\GLAM}

\author{Harold~Y.~Hwang}
\affiliation{\StanfordAP}
\affiliation{\SIMES}

\author{Colin~Ophus}
\thanks{\href{mailto:cophus@stanford.edu}{cophus@stanford.edu} (C.O.) and \href{mailto:alfredz@stanford.edu}{alfredz@stanford.edu} (A.Z.)}
\affiliation{\StanfordMSE}
\affiliation{\SIMES}

\author{Alfred~Zong}
\thanks{\href{mailto:cophus@stanford.edu}{cophus@stanford.edu} (C.O.) and \href{mailto:alfredz@stanford.edu}{alfredz@stanford.edu} (A.Z.)}
\affiliation{\StanfordAP}
\affiliation{\SIMES}
\affiliation{\StanfordP}

\date{\today}

\begin{abstract}
Freestanding thin films of quantum materials naturally develop sub-micrometer, nonuniform strain fields that strongly affect their electronic, magnetic, and structural properties in both equilibrium and nonequilibrium conditions. However, current methods to quantitatively resolve these mesoscopic features are primarily restricted to scanning probes, making it challenging for dynamical measurements such as single-shot imaging and femtosecond microscopy. Here, we present a new computational framework which we denote as \textit{bend contour electron tomography} (BCET), which efficiently converts bend contours in transmission electron microscope images into quantitative two-dimensional maps of strain and topography. By iteratively minimizing a designed loss function between experimental and simulated bend contour images, BCET retrieves both surface morphology and in-plane strain tensor fields without requiring scanning or diffraction mapping. We applied BCET to freestanding SrTiO$_3$ thin films, demonstrating the successful reconstruction of the local strain distribution and curvature field with high fidelity. Our approach provides a quantitative framework for characterizing mesoscale structures in freestanding films using wide-field imaging, opening new avenues to investigate how spatial inhomogeneity governs phase transitions and nonequilibrium dynamics in two-dimensional quantum materials.

\end{abstract}

\maketitle

\section*{I\lowercase{ntroduction}}

Thin films provide a versatile platform to explore reduced dimensionality, where out-of-plane boundary conditions modify electronic and vibrational properties. For example, reducing film thickness can drive metal–insulator transitions \cite{scherwitzl_MetalInsulator_2011} and modify superconducting properties \cite{guo_superconductivity_2004}, while also giving rise to sub- or super-sonic acoustic modes not accessible in a bulk crystal \cite{lamb_waves_1917, viktorov_Rayleigh_1967, dekorsy_Coherent_2000}. Furthermore, reduced dimensionality imposes fundamental thermodynamic constraints on thin films: long-wavelength fluctuations destabilize ideal crystals strictly confined in two dimensions \cite{mermin_Crystalline_1968}, but real thin films remain stable because anharmonic coupling between flexural and in-plane modes renormalizes the effective bending rigidity and elastic response, leading to scale-dependent roughness and mesoscopic out-of-plane distortions. In reality, finite-size membranes are often subjected to residual stress, clamping, and disorder, which manifest as mesoscopic out-of-plane wrinkles and associated strain heterogeneity \cite{nelson_Fluctuations_1987, aronovitz_Fluctuations_1988, j.bowick_Statistical_2001, fasolino_Intrinsic_2007}.

The intrinsic mesoscale wrinkles generate nonuniform strain and curvature fields that couple to collective degrees of freedom in freestanding thin films. In thermal equilibrium, gradients in a strain field can bias ferroic domain structures \cite{harbola_strain_2021, lukashev_Flexomagnetic_2010, jiao_flexoelectricitystabilized_2023}, and periodically modulated strains can even produce a pseudo-magnetic field and lead to Landau quantization \cite{milovanovic_band_2020, mao_evidence_2020}. Following ultrafast laser excitation, such nonuniformity can seed  spatially heterogeneous phase transition pathways, leading to distinct outcomes in a nonequilibrium state \cite{ahn_Xray_2022, zhu_Mesoscopic_2016}. Designing wrinkles and related mesoscale structures hence becomes a central part of thin film engineering, where patterned substrates, pre-strain, or kirigami are leveraged to create new functional properties \cite{kim_strain_2023, so_polarization_2021, daveau_spectral_2020, blees_graphene_2015, miskin_graphenebased_2018}.

These developments are underpinned by the advancement in microscopy to quantitatively measure the mesoscale strain and heterogeneity in freestanding thin films. Scanning probes have been a major workhorse: scanning electron or x-ray nano-diffraction microscopy can map strain tensor or topography fields by analyzing position-dependent diffraction patterns \cite{chahine_Strain_2015, savitzky_py4dstem_2021, munshi_disentangling_2022, ribet_Multiangle_2025, mireles_Strain_2026}, while tip-based atomic force microscopy and scanning tunneling microscopy can also provide atomic-resolution topographic maps of freestanding films and membranes \cite{lee_Measurement_2008, neek-amal_Thermal_2014}. An alternative route leverages interferometry using a highly coherent beam, where methods such as dark-field electron holography and coherent x-ray diffraction imaging have demonstrated quantitative strain mapping with high precision \cite{robinson_Coherent_2009,beche_Dark_2011}. 

Although these approaches provide powerful, quantitative access to morphology and strain in thin films, they are often constrained to quasi-static or steady-state measurements. To investigate how mesoscale heterogeneity emerges, evolves, and couples to structural and electronic order out of equilibrium, one needs an efficient microscopy approach that is compatible with \textit{operando} or ultrafast pump-probe measurements. Scanning-based methods often suffer from uncontrolled drift and prolonged acquisition time in pump-probe measurements \cite{nakamura_Visualizing_2022}, while holography-based measurements require stringent beam conditions. In this regard, wide-field transmission electron microscopy (TEM) offers a promising route to probe nonequilibrium states. At the pico- to nanosecond timescales, ultrafast TEM has demonstrated the ability to probe photoinduced structural dynamics like propagating elastic waves, revealing rich mesoscale thin film physics following laser excitation \cite{zhang_Observation_2019, zhang_Imaging_2021, zong_Spinmediated_2023}. At the micro- to millisecond timescales, \textit{in-situ} TEM enables direct visualization of film evolution under thermal, electrical, or mechanical stimuli, providing real-time access to dynamical structural processes across mesoscale fields of view \cite{minor_New_2006, yu_Situ_2015, curtis_Ultrathin_2026}. Yet, strain contrast in wide-field TEM remains qualitative: Intensity is a nonlinear projection of local strain and tilt, without a general protocol to extract quantitative topography or strain tensor fields. 

To address this gap, we developed bend contour electron tomography (BCET), a computational framework that provides quantitative maps of strain and morphology based on wide-field TEM images of single-crystalline thin films. We use the term \emph{tomography} to capture the general notion of object reconstruction by combining multiple measurements acquired under different projection conditions, most commonly by varying the sample tilt angles as we will show in this work. BCET extracts the latent structural information encoded in bend contour contrast, converting qualitative intensity variations into spatially-resolved in-plane strain tensor and surface height profiles. By leveraging full-field acquisition, this approach enables fast mapping without scan distortions in both static and dynamical experiments. BCET thus establishes a pathway to interrogate how strain inhomogeneity evolves and couples to electronic, vibrational, and ferroic phenomena under equilibrium and far-from-equilibrium conditions, transforming wide-field electron microscopy into a quantitative structural probe at the mesoscale. 

\section*{B\lowercase{end contours}}

\begin{figure}[hbt!]
    \includegraphics[width=1\linewidth]{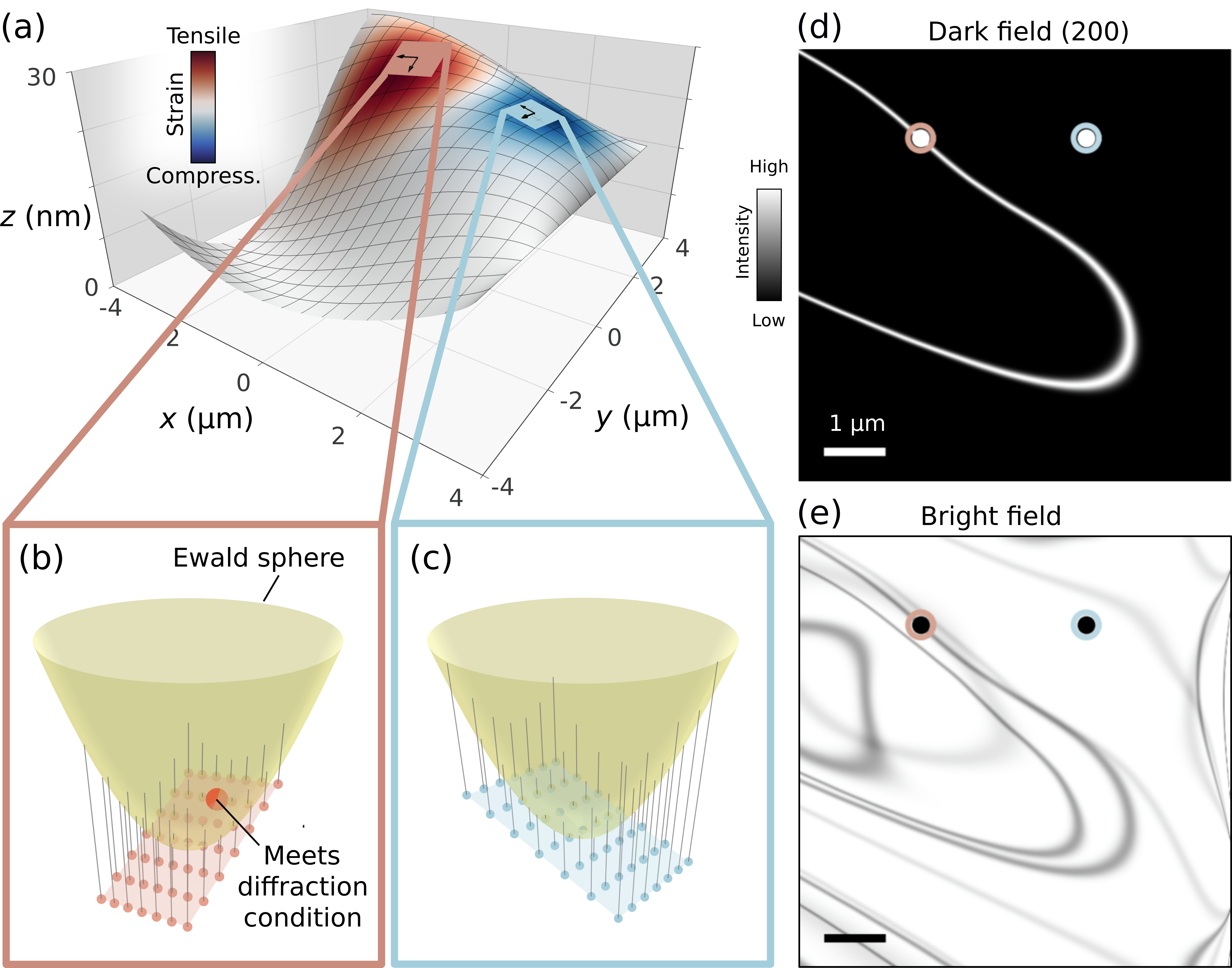}
    \caption{\textbf{Principles of bend contours.} (a)~An example freestanding thin film of a cubic SrTiO$_3$ lattice that features two regions with compressive (blue) and tensile (red) strains. Tangent planes are drawn at the two respective locations on the film surface. (b,~c)~Schematic of the local reciprocal lattice at each point. Black lines extending to the Ewald sphere (yellow) represent the deviation from meeting the Bragg condition; the graphics are vertically stretched for clarity. For the blue region, no diffraction condition is met as none of the reciprocal lattice points reside on the Ewald sphere. In the pink region, the (200) peak (red dot) resides on the Ewald sphere, leading to locally scattered electrons. (d)~Simulated (200) dark-field image computed by the forward model based on the film morphology and strain in (a). It features the (200) bend contour intersecting with the pink region but not the blue region. (e)~Simulated bright-field image that include all possible bend contours given the film curvature. Because no diffraction occurs at the blue point, no bend contour intersects with it in real space.}
    \label{fig:1}
\end{figure}

Formulating BCET relies on understanding the primary contrast mechanism in low-magnification TEM of single-crystalline films, known as bend contours \cite{williams_Transmission_2009}. Historically they have played an important role in crystallography, finding uses in the characterization of crystal orientation, polarity, symmetry, morphology, and topological defects \cite{buxton_Symmetry_1976, spiecker_Determination_2002, spiecker_Quantitative_2003, shannon_Bend_1979, bolotov_Electron_1982, bolotov_Electron_1982a, pham_Microscopic_2025}. However, quantitative analysis of bend contours remains formidable due to an astronomical number of operations involved in the computation, which only becomes viable following recent breakthroughs in parallel computing.

The mechanism behind bend contour contrast is illustrated through an example wrinkled thin film in Fig.~\ref{fig:1}(a), which features two local patches and their corresponding reciprocal lattices [Fig.~\ref{fig:1}(b,c)]. At the pink patch in Fig.~\ref{fig:1}(a), the local tilt and strain conditions meet the diffraction condition for the $(200)$ Bragg peak, so electrons are diffracted to the corresponding point on the $(200)$ dark field image [Fig.~\ref{fig:1}(d)]; consequently, intensity is removed from that part of the bright-field image [Fig.~\ref{fig:1}(e)]. At the blue patch, no reciprocal lattice vector meets the diffraction condition, resulting in no bend contours intersecting that point. Tilting the sample will globally rotate each local region, causing new points to meet the diffraction condition and bend contours to translate and distort along the surface.

More formally, bend contours trace loci on the film where a particular reciprocal lattice vector satisfies the elastic scattering condition, $\mathbf{k}_f - \mathbf{k}_i=\mathbf{G}_{hkl}$, where $\mathbf{k}_i$ and $\mathbf{k}_f$ are the incoming and outgoing wavevectors, and $\mathbf{G}_{hkl}$ is the reciprocal lattice vector for the Miller indices, $(h,k,l)$. On the loci, the bright-field intensity $I^{\text{bf}}$ is reduced due to the redistribution of intensity to diffracted beams. Bend contour contrast is therefore sensitive to local tilt, which rotates the reciprocal peaks into the Ewald sphere, and local strain, which contracts, expands, or shears the reciprocal lattice. For a complete mathematical description of bend contour contrast, see Sec.~\ref{supp:bc_math} of the Supplemental Material.

\section*{A\lowercase{lgorithmic overview}}

To convert bend contour contrast into quantitative measures of strain and tilt, we cast BCET as a physics-constrained inverse problem. We first develop a forward model that predicts bright-field bend contour images at different tilt angles from a parameterized description of the surface morphology and in-plane strain tensor fields. The reconstruction is then posed as the inversion of this forward model: we search for the surface and strain fields whose simulated contrast most closely matches the experimental observations. To this end, we introduce a loss functional that compares measured and simulated images over all tilt conditions and iteratively optimize the structural parameters. The resulting converged solution provides a self-consistent estimate of the surface topography and in-plane strain, translating \textit{qualitative} bend contour images into \textit{quantitative} structural maps.

\section*{F\lowercase{orward model from film topography and strain to bend contours}}

The forward model predicts bright- and dark-field contrast from a parameterized surface geometry, $\mathbf{R}(u,v)$, where $u,v\in [0,1]$ are parametric coordinates. We start by computing the local surface tangent plane and normal vector of $\mathbf{R}(u,v)$, which define the crystal orientation at each point. For a given set of Miller indices, $(hkl)$, we then compute the corresponding dark-field intensity $I^{\text{df}}_{hkl}(u,v)$ by evaluating the kinematic diffraction response as a function of the position-dependent minimal distance to the Ewald sphere determined by the local lattice orientation and strain. In this formulation, spatial variations in surface tilt modify reciprocal lattice orientation, while strain modifies reciprocal lattice spacing, jointly modulating the Bragg conditions and producing spatially dependent bend contour contrast. The bright-field intensity $I^{\text{bf}}(u,v)$ is obtained by enforcing conservation of total intensity, such that contrast arises from the redistribution of intensity between transmitted and diffracted beams. This deterministic forward model defines a mapping from a continuous surface morphology [Fig.~\ref{fig:1}(a)] to simulated dark- and bright-field images [Fig.~\ref{fig:1}(d,e)]. The full mathematical details of the forward model along with considerations of multiple or inelastic scattering, beam coherence, and thickness inhomogeneity are provided in Sec.~\ref{supp:forward_model} of the Supplemental Material.

\section*{I\lowercase{nitial estimate via a direct solver}}

\begin{figure}[t!]
    \centering
    \includegraphics[width=1\linewidth]{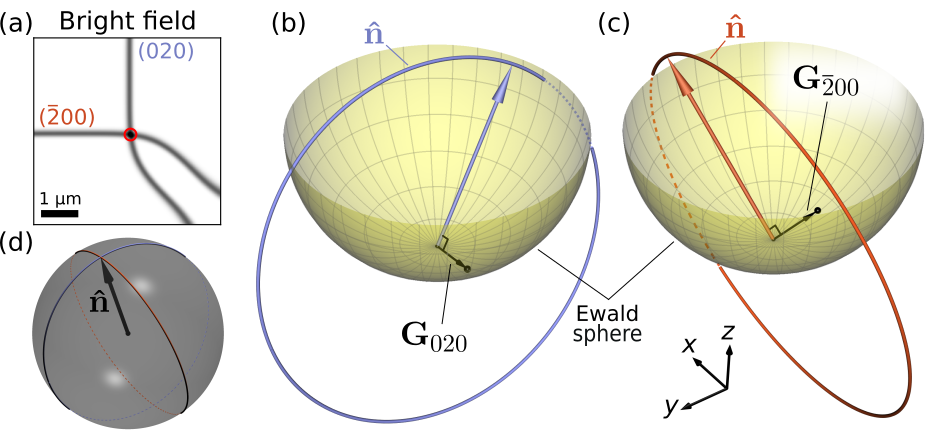}
    \caption{\textbf{Surface normal constrained by the bend contours.} (a)~Simulated bright-field electron micrograph with two intersecting bend contours: $(\overline{2}00)$ and $(020)$. (b)~Illustration of the constraints on the surface normal vector $\hat{\mathbf{n}}$ at positions along the $(020)$ contour. First, $\mathbf{G}_{020}$ is constrained to the Ewald sphere. Second, $\hat{\mathbf{n}}$ must be perpendicular to $\mathbf{G}_{020}$ in this cubic lattice example. Hence, $\hat{\mathbf{n}}$ must fall on to a point prescribed by the blue circle. (c)~Schematic for the corresponding normal vector constraint at all locations on the $(\overline{2}00)$ contour. (d)~At the intersection point of the two bend contours [red circle in (a)], combining constraints from (b) and (c) uniquely determines $\hat{\mathbf{n}}$ up to a sign.}
    \label{fig:2}
\end{figure}

Before we perform iterative optimization for the inverse solver, we need a nearby solution to arrive at a successful reconstruction. For this purpose, we developed a direct solver to supply the initial guess. For simplicity, this algorithm assumes strain-free conditions, with uniform lattice constants ($a$, $b$, $c$) and lattice angles ($\alpha$, $\beta$, $\gamma$) across the film (see Fig.~\ref{fig:forward_sim}(a) of the Supplemental Materials). This direct solver formulates the presence of bend contours as mathematical constraints on the surface topography, and solves for a surface that meets these combined constraints using a least-squares approach. 

The direct solver was inspired by the observation that bend contour contrast originates from the local reciprocal lattice meeting the diffraction condition due to tilt and strain. For simplicity, here we illustrate the geometric constraints imposed by the observation of bend contours for a cubic lattice in Fig.~\ref{fig:2}; for a general treatment of arbitrary three-dimensional crystalline lattices, see Sec.~\ref{supp:general_direct_constraints} of the Supplemental Material. Two nonparallel bend contours, $(\overline{2}00)$ and $(020)$, are visible in the bright-field image [Fig.~\ref{fig:2}(a)]. At each point on the contour, the surface must be tilted such that the reciprocal lattice vector, $\mathbf{G}_{hkl}$, intersects the Ewald sphere. The constraint is most easily illustrated for an in-plane vector, $\mathbf{G}_{hk0} = h\mathbf{a}^* + k\mathbf{b}^*$, where $\mathbf{a}^*$ and $\mathbf{b}^*$ are the reciprocal lattice unit vectors. In this case, the surface normal vector in real space, $\hat{\mathbf{n}} \parallel (\mathbf{a} \times \mathbf{b})$, lies perpendicular to the $\mathbf{a}, \mathbf{b}$ plane, and therefore it is perpendicular to $\mathbf{G}_{hk0}$ [Fig.~\ref{fig:2}(b,c)]. This restricts the local surface tilt along a bend contour, but not fully: the normal vector $\hat{\mathbf{n}}$ can still be rotated arbitrarily around $\mathbf{G}_{hk0}$ while continuing to meet the diffraction condition. The possible normal vectors given the diffraction constraints are shown in Fig.~\ref{fig:2}(b,c) as the colored circles. With multiple nonparallel bend contour observations at a single location, we can fully determine the normal vector $\hat{\mathbf{n}}$ [Fig.~\ref{fig:2}(d)]. However, with only a single bright-field image, the topography of the surface is under-defined in general. To determine the normal vector at each position, we require at least two nonparallel bend contours to traverse over the field of view. This condition can be achieved by tilting the film. If more bend contours are visible, their constraints can be incorporated into a global fit to improve the accuracy of the surface-normal reconstruction.

We formulate the normal-vector constraints as per-pixel height conditions along the bend contours and stack them into a matrix equation, which we solve using least-squares to recover the surface topography. In practice, a regularization term is also added to help enforce smoothness and continuity. For a more detailed mathematical description, results, and validation of the direct solver---including both cubic and lower-symmetry lattices---see Secs.~\ref{supp:general_direct_constraints} and \ref{supp:low_symmetry_recon} of the Supplemental Material.

\section*{I\lowercase{terative reconstruction of topography and strain tensor maps}}

While the direct solver provides an initial estimate of surface geometry, it neglects strain and depends on experimental parameters that are hard to determine. Inspired by the iterative algorithms developed for conventional electron tomography and inverse imaging problems \cite{ren_multiple_2020, pham_Accurate_2023, mccray_accelerating_2025, ding_Threedimensional_2022}, we overcome these limitations by employing an iterative inverse reconstruction framework, which relaxes the strain-free assumptions and refines experimental parameters.

The goal of this algorithm is to iteratively solve for an unknown parameter vector, $\mathbf{x}$, that encodes the sample surface topography and strain. The surface geometry is represented by a mesh of control points defining a B\'ezier surface \cite{piegl_Bspline_1997}, which provides a smooth and compact parameterization suitable for gradient-based optimization (see Sec.~\ref{supp:Bezier} of the Supplemental Material). Given a trial parameter vector $\mathbf{x}$ and a tilt angle $i$, we compute simulated bright-field bend contour images $I^{\text{bf}}(\mathbf{x}, i)$ and compare them to the corresponding experimental images $I^{\text{bf}}_0(i)$. We quantify this comparison using a loss function

\begin{equation}\label{equation:cost_function}
    \epsilon(\mathbf{x}) = \sum_i C\left[I^{\text{bf}}_{0}(i), I^{\text{bf}}(\mathbf{x}, i)\right],
\end{equation}
where $C$ is an image similarity metric, such as the mean squared error (MSE). Minimizing $\epsilon(\mathbf{x})$ with respect to $\mathbf{x}$ yields a best-fit estimate of the surface topography and strain. The optimization is performed using a gradient-based approach implemented in \textit{PyTorch} \cite{paszke_pytorch_2019a}, with parameters updated using the \textit{Adam} optimizer \cite{kingma_adam_2015}. Because our forward model is designed in a way that it is fully differentiable with respect to $\mathbf{x}$, gradients of the simulated intensities can be computed efficiently using automatic differentiation (see Sec.~\ref{supp:optimization_loop} of the Supplemental Material).

For the highly non-convex loss landscape, we employ a pyramidal optimization strategy to improve convergence, as demonstrated in scientific computing \cite{saad_13_2003, thevenaz_pyramid_1998}, machine learning \cite{chen_graph_2022}, and tomography \cite{perelli_stochastic_2024}. In this approach, the reconstruction is first performed using a low-resolution B\'ezier surface with a small number of control points, allowing large-wavelength features to be captured efficiently. The optimized surface is then interpolated onto progressively finer control point meshes and re-optimized (see Sec.~\ref{supp:Bezier} of the Supplemental Material). This multiscale strategy significantly improves robustness compared to directly optimizing a high-dimensional parameterization.

\section*{R\lowercase{econstruction of a freestanding oxide perovskite thin film}}

\begin{figure*}[hbt!]
    \centering
    \includegraphics[width=1\linewidth]{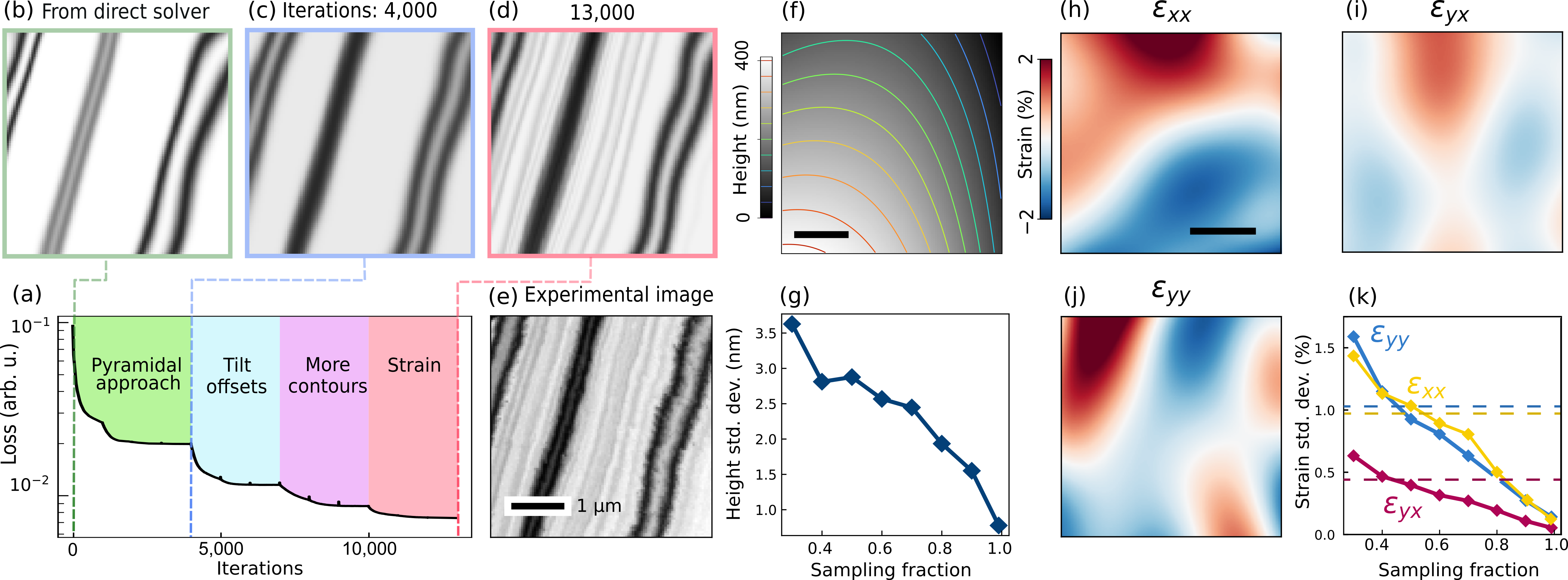}
    \caption{\textbf{Iterative reconstruction of a SrTiO$_\text{3}$ thin film.} (a)~The loss function in Eq.~\eqref{equation:cost_function} over different iterations. Colored regions refer to different parts of the multiphase optimization procedure; see main text and Sec.~\ref{exp_recon_details} of the Supplemental Material for details. (b--d)~Simulated bright-field micrographs at zero-tilt throughout the multiphase optimization: (b)~initial guess from the direct solver, (c)~after the pyramidal optimization on the B\'ezier control points, and (d)~the final converged solution. (e)~Experimental image at zero tilt; data at other tilt angles are shown in Fig.~\ref{fig:experimental_images_example} of the Supplemental Material. High spatial frequency deviations between the optimized [panel~(d)] and experimental images could arise from nanoscale strain fields or flexural features with nanometer-scale wavelengths that cannot be captured by $20 \times 20$ B\'ezier control points used in the iterative optimization. (f)~Reconstructed height map with equal height contours. (h--j)~Reconstructed strain tensor maps computed using a flat reference configuration such that the globally averaged strain is zero for each strain channel. All scale bars in (b--f) and (h--j) are \SI{1}{\micro\metre}. 
    (g,k)~Cross validation to assess the precision of the algorithm, where the standard deviations of reconstructed height (g) and strains (k) are computed for 40 reconstructions for each sampling fraction of the experimental images. Horizontal dashed lines in (k) are the average reconstructed strain using all experimental data; the average height for (g) is 73~nm, which is far outside the plotting window. When the standard deviation falls well below the average, the reconstruction is stable and converges to a consistent solution.}
    \label{fig:3}
\end{figure*}

For experimental validation of BCET, we selected strontium titanate (SrTiO$_3$), a perovskite oxide exhibiting incipient ferroelectric correlations. Although SrTiO$_3$ avoids a ferroelectric phase transition at low temperature due to quantum fluctuations \cite{muller_SrTiO3_1979}, it lies in close proximity to ferroelectric order. Perturbations including defects, interfaces, or reduced dimensionality can offset the energy landscape and stabilize polar distortions \cite{schlom_Strain_2007, xu_Straininduced_2020, wordenweber_Induced_2007, jang_Ferroelectricity_2010}. In thin films, strong flexoelectric coupling further links strain gradients to polarization, enabling curvature-induced polar responses at the nanoscale \cite{shang_Synthesizing_2024}. Together with recent advances in freestanding oxide film synthesis \cite{lu_Synthesis_2016}, these properties make SrTiO$_3$ an ideal platform for applying BCET to spatially resolve mesoscale strain fields.

We acquired a tilt series of electron micrographs on a room temperature 30-nm-thick SrTiO$_3$ freestanding film using a 300~keV transmission electron microscope. A total of 40 images were recorded, rotating about two perpendicular tilt axes in a range of $\pm5^\circ$ for each axis. A representative image at zero tilt is shown in Fig.~\ref{fig:3}(e); data at other tilt angles are shown in Fig.~\ref{fig:experimental_images_example} of the Supplemental Material. Tilt series images were registered relative to each other using the underlying Cu mesh boundary. Bend contours in each bright-field image were manually masked by their diffraction order to generate virtual dark field images for constraining the direct solver, a manual step that took around four to six hours for the presented dataset but can potentially be automated via an artificial intelligence-assisted workflow (see details in Sec.~\ref{supp:dark_field_labelling} of the Supplemental Material). Based on the virtual dark field images, 298,572 constraints on the surface height map were formulated and supplied to the direct solver to reconstruct the surface topography. The corresponding bright-field image of the resulting surface is shown in Fig.~\ref{fig:3}(b), which is the initial guess for the subsequent iterative optimization.

For the iterative reconstruction, we employed a multi-phase strategy in which the number of free variables in $\mathbf{x}$ was increased progressively. In each phase labeled by different colors in the loss curve [Fig.~\ref{fig:3}(a)], additional optimization parameters were introduced while previously optimized parameters were allowed to relax further. This hierarchical approach stabilizes the optimization and improves convergence by first fitting the most strongly constrained degrees of freedom before refining higher-order corrections. In the first stage, the pyramidal approach is utilized: increasing the control point number every 1,000 iterations starting from 10 $\times$ 10 to 20 $\times$ 20 after 4,000 iterations [Fig.~\ref{fig:3}(c)]; see Sec.~\ref{supp:control_pt_num_determine} of the Supplemental Material for details on how we determined the best control point mesh size. In the blue, purple, and red regions of Fig.~\ref{fig:3}(a), tilt offsets, extra bend contour orders, and strain, respectively, are sequentially added to $\mathbf{x}$ for optimization. Further details of the multiphase optimization procedure are provided in Sec.~\ref{exp_recon_details} of the Supplemental Material. After 13,000 total iterations, the reconstruction converges to a stable solution with quantitative agreement between simulated and experimental images [Fig.~\ref{fig:3}(d,e)]. Corresponding height and strain components of the converged solution are shown in Fig.~\ref{fig:3}(f,h--j). The reconstructed strain magnitudes are well above the strain sensitivity of approximately 0.1\% granted by our experimental noise conditions (see details in Sec.~\ref{supp:strain_sensitivity} of the Supplemental Material). The effective resolution of this reconstruction is approximately 200~nm, estimated by the density of control points over the field of view. In total, the convergence of 1,279 parameters in $\mathbf{x}$ was reached after less than 10~minutes using a single NVIDIA L40 graphics processing unit.

The mesoscale strain fields recovered here have direct physical consequences for SrTiO$_3$. Its paraelectric-to-ferroelectric transition temperature rises steeply above zero with strain and can even reach room temperature or above at few percent level strains \cite{haeni_Roomtemperature_2004,xu_Straininduced_2020,li_Classicaltoquantum_2025}. The $\sim$1--2\% local strains revealed by BCET [Fig.~\ref{fig:3}(h--j)] are therefore comparable in magnitude to values shown to induce ferroelectricity in previous studies \cite{xu_Straininduced_2020, li_Classicaltoquantum_2025}. Because this strain field is spatially heterogeneous, the film does not occupy a single point on the strain--temperature phase diagram but rather a continuous distribution of local states, with different regions of the film predicted to cross the ferroelectric phase boundary at different temperatures. Quantifying and visualizing this mesoscale strain landscape are precisely the class of questions BCET is designed to address. The resulting strain maps provide a quantitative structural foundation for interpreting the spatial organization of strain-driven phenomena, such as ferroelectric \cite{li_Classicaltoquantum_2025}, magnetic \cite{cenker_Reversible_2022}, or even multiferroic orders \cite{lee_EpitaxialStrainInduced_2010, sun_Straininduced_2025}.

\section*{V\lowercase{alidation}}

We validated the BCET algorithm along three axes: (i)~identifiability analysis, which establishes that the inverse problem admits a unique solution given sufficient tilt data, (ii)~subsampling cross-validation analysis, which quantifies the precision and stability of the reconstruction, and (iii)~simulated reconstruction analysis under varying noise conditions, which evaluates accuracy.

A central concern for any inverse problem in tomography is whether the reconstructed solution is unique. In the case of BCET, local tilt and in-plane strain can be entangled as both factors enter the diffraction condition in a coupled fashion. A single bright-field image would underconstrain the inverse problem, in the same way that a single projection underconstrains a three-dimensional object in conventional tomography. To break the degeneracy between spurious and true solutions, we need to add independent constraints by tilting the sample, which can eventually yield a unique solution in an overdetermined inverse problem.

To validate the uniqueness of BCET, we relied on a general concept in optimization. For an underconstrained loss function, many local minima exist. Different starting parameters in the optimization procedure will result in uncorrelated and spurious solutions with a wide distribution of strain and height errors from the ground truth. On the other hand, if the solution is unique by including appropriate constraints, the loss function will fall into a single minimum. In this case, the algorithm will yield a narrow distribution of loss, strain errors, and height errors for different starting conditions. As detailed in Sec.~\ref{supp:uniqueness} of the Supplemental Material, with a single tilt image, the converged loss, height, and strain errors are indeed broadly distributed, showing signs of a non-unique problem with many spurious solutions. As more tilt angles are included as input to the BCET algorithm, these distributions collapse to a tight, low-error cluster, marking the transition to a unique reconstruction. This transition is controlled jointly by the number of tilt images and the tilt range: what matters is that several non-parallel bend contours sweep across the full field of view while moving in small increments between adjacent tilts, so that each region of the freestanding film accumulates a sufficient number of constraints. As we show in Sec.~\ref{supp:uniqueness} of the Supplemental Material, our experimental dataset (40~images over a $\pm5^\circ$ tilt range) lies well inside this uniquely identifiable regime, and the same analysis indicates that comparably well-constrained reconstructions are achievable with far fewer images under an appropriately chosen tilt sampling.

In our cross-validation study to assess precision and stability of the BCET algorithm, we repeatedly run the final step of the multiphase reconstruction on random subsets of the experimental images. For each sampling fraction, we generate 40 reconstructions, each based on a newly selected data subset. If the algorithm converges to a consistent solution, the variation among the 40 reconstructions should be small compared to the mean value. Motivated by this intuition, at each sampling fraction, we computed the standard deviation across the reconstructed height and the three independent components of the strain tensors, shown in Fig.~\ref{fig:3}(g,k). After a sampling fraction of 0.5, the standard deviation among the 40 reconstructions of each strain tensor falls below the mean [see horizontal dashed lines in Fig.~\ref{fig:3}(k)]; for the height, the variation is way below the mean of 73~nm even at the smallest sampling fraction of 0.3. These observations indicate that the reconstruction is stable and converges to a consistent solution, independent of the specific subset of input images. More details of the precision validation can be found in Sec.~\ref{supp:precision} of the Supplemental Material.

\begin{figure}[t!]
    \centering
    \includegraphics[width=1\linewidth]{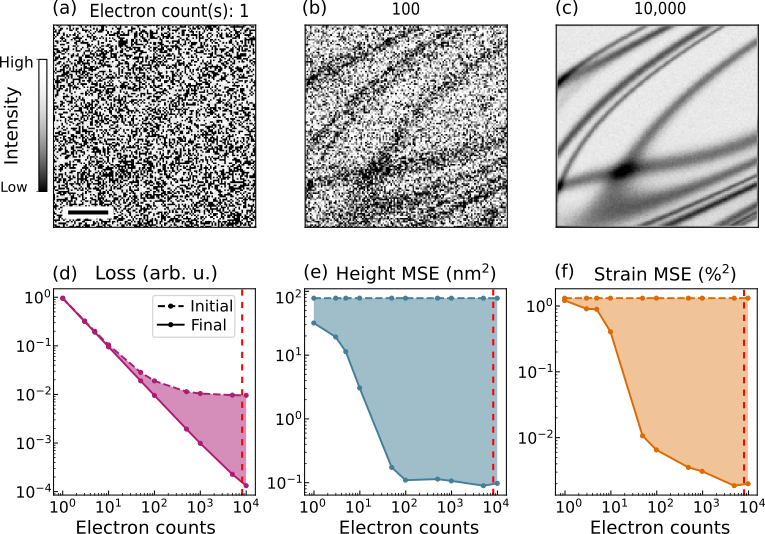}
    \caption{\textbf{Assessment of reconstruction accuracy at varying noise levels}. (a--c)~Representative simulated electron micrographs with Poisson noise generated using 1, 100, and 10,000 electron counts per pixel, respectively. Scale bar:~\SI{1}{\micro\metre}. (d--f)~Electron count-dependent MSE compared to the ground truth for the initial and final loss, height, and strain. Vertical dashed lines denote the experimental noise condition, estimated using a screen current of 1.38~nA for 1~s acquisition using a detector with $1,024\times1,024$ pixels.}
    \label{fig:4}
\end{figure}

To validate the accuracy of BCET, we performed reconstructions on simulated bend contour images with different noise levels and compared the result to the ground truth surface (see details in Sec.~\ref{supp:accuracy} of the Supplemental Material). Poisson-distributed shot noise, originating from the discrete nature of electron counting, was added to each pixel of the bright-field images. Dark current and read noise were omitted because they are insignificant relative to the shot noise under standard TEM conditions \cite{williams_Transmission_2009}. We varied the electron counts for the simulated shot noise from 1 to 10,000 at ten different values and used the iterative method to reconstruct the surface for each noise level. Figure~\ref{fig:4}(a--c) shows the simulated electron micrographs with different Poisson noise levels defined by the electron counts per pixel. For each noise level, we perform an iterative reconstruction and report the count-dependent initial and final loss, height MSE, and strain MSE in Fig.~\ref{fig:4}(d--f). The converged loss value is highly dependent on the noise level at all electron counts, but the height MSE plateaus near 100 electron counts at 0.1~nm$^2$. The strain MSE reaches a relatively small value ($<0.01\%^2$) at 100 electron counts, but continues to improve with electron counts. This behavior indicates that the algorithm maintains quantitative reconstruction accuracy down to approximately 100 electrons per pixel [Fig.~\ref{fig:4}(b)]. This level of electron count is well below the average experimental count of 8,200 electrons/pixel [see the vertical dashed lines in Fig.~\ref{fig:4}(d--f)]. These results demonstrate that BCET accurately recovers both surface topography and strain fields under realistic noise conditions.

Although we performed the experimental and validation studies on the high-symmetry cubic crystal of SrTiO$_3$, reconstructing materials with a lower-symmetry space group does not introduce additional computational complexity or degraded performance. To demonstrate this capability, we performed reconstruction on a synthetic dataset of a monoclinic film of Ga$_2$O$_3$. Details can be found in Sec.~\ref{supp:low_symmetry_recon} of the Supplemental Material.

\section*{D\lowercase{iscussion}}

BCET relies on full-field acquisition rather than probe scanning, and therefore is uniquely suited for probing mesoscale structural heterogeneity with a large field of view up to hundreds of micrometers. This regime is difficult for typical scanning-based microscopy such as atomic force microscopy and scanning tunneling microscopy. Although BCET does not provide atomic resolution, it fills a critical gap between nanoscale probes and macroscopic optical microscopy methods. Importantly, BCET is uniquely suited for single-shot, ultrafast, and \textit{in-situ} measurements for several reasons. First, eliminating the need for probe scanning minimizes the adverse effect of sample drift, making long-time data acquisition more robust. Second, BCET can be implemented on virtually any conventional TEM setup and it does not need a specialized tomography sample holder. Third, the relatively small amount of data needed for reconstruction compared to scanning nanobeam electron diffraction and other data-intensive methods makes it possible for efficient or even real-time data processing. Furthermore, the computational framework introduced here can be straightforwardly extended to accommodate a large number of additional parameters in $\mathbf{x}$, especially in light of the rapid advances and widespread availability of parallel computing today. These attributes establish BCET as a broadly accessible approach for resolving complex structural landscapes in crystalline thin films, which also provides a versatile platform for studying flexo-ferroic coupling, thin film acoustic excitations, and emergent functionality arising from strain gradients in freestanding membranes \cite{jiao_flexoelectricitystabilized_2023, viktorov_Rayleigh_1967, schlom_Strain_2007}. 

Several technical extensions can further expand the scope and impact of BCET. By measuring bend contours beyond the first Laue zone, the same framework can be used to recover the spatial map of out-of-plane strains that typically have the fastest response following photoexcitation by a femtosecond laser pulse \cite{cremons_Defectmediated_2017}. We demonstrate this possibility using a synthetic dataset in Sec.~\ref{supp:c_lattice_recon} of the Supplemental Material. Incorporating film surface models that fit multi-domain crystalline regions using a piecewise continuous function would enable quantitative mapping of crystallographically heterogeneous systems using BCET. This capability would allow direct visualization of domain switching or even topological defect motion, which play a central role in nonequilibrium phase transitions \cite{cheng_Ultrafast_2024a}. Furthermore, applying BCET to reconstruct more intricate surface geometries, such as folded sheets or patterned membranes exemplified in kirigami structures \cite{blees_graphene_2015}, would make it possible to directly visualize the strong strain localization and pronounced curvature that develop along folded creases or cut features \cite{shyu_Kirigami_2015}. These capabilities would establish BCET as a powerful tool for resolving complex mesoscale structural landscapes that remain difficult to quantify with existing techniques.

By reconstructing surface topography and the full in-plane strain tensor, BCET enables direct visualization of strain gradients and curvature that underlie higher-order electro- and magneto-mechanical couplings \cite{schlom_Strain_2007,jiao_flexoelectricitystabilized_2023}. These effects are particularly sensitive to mesoscale structural heterogeneity, including bending, wrinkling, and local curvature variations common in freestanding thin films \cite{j.bowick_Statistical_2001, fasolino_Intrinsic_2007,makushko_flexomagnetism_2022,jani_spatially_2024}. BCET therefore provides a quantitative structural framework for \textit{operando} investigation of emergent phenomena in low-dimensional materials driven by strain and nontrivial topography. Our work further illustrates how combining a widely accessible full-field imaging technique with modern parallel computational architectures offers a powerful strategy for quantitative microscopy.
   
\vspace{0.55 cm}
\noindent\textit{Acknowledgments} --- We thank helpful discussions with Cedric~Lim, Patrick~Liu, and Jared~Maxson. We thank Andrew~Barnum and Pinaki~Mukherjee for assistance with the TEM measurements. We thank Jiarui~Li for SrTiO$_3$ characterization. C.O. acknowledges funding support from the DOE Office of Science FWP-101256, Enabling Science for Transformative Energy-Efficient Microelectronics. H.Y.H and M.H. acknowledge funding from the U.S. Department of Energy, Office of Basic Energy Sciences, Division of Materials Sciences and Engineering, under contract No.~DE-AC02-76SF00515. A.Z. and H.G.B. acknowledge the support from the U.S. Department of Energy, Office of Basic Energy Sciences under award No.~DE-SC0026202. Part of this work was performed at nano@stanford (RRID:SCR\_026695), and we thank the C-ShaRP Voucher Program for supporting the facility access.

\clearpage
\onecolumngrid

\renewcommand{\theequation}{S\arabic{equation}}
\renewcommand{\thefigure}{S\arabic{figure}}
\renewcommand{\thetable}{S\arabic{table}}
\renewcommand{\thesection}{S\arabic{section}}
\renewcommand{\thesubsection}{S\arabic{section}.\arabic{subsection}}
\renewcommand{\theHequation}{S\arabic{equation}}
\renewcommand{\theHfigure}{S\arabic{figure}}
\renewcommand{\theHtable}{S\arabic{table}}
\renewcommand{\theHsection}{S\arabic{section}}
\renewcommand{\theHsubsection}{S\arabic{section}.\arabic{subsection}}
\setcounter{equation}{0}
\setcounter{figure}{0}
\setcounter{table}{0}
\setcounter{section}{0}

\twocolumngrid

\onecolumngrid
\begin{center}

\textbf{\large Supplemental Material to ``Bend Contour Electron Tomography (BCET): \\ Quantitative strain and topography mapping''}\\[6pt]
\vspace{0.25 cm}
Henry~G.~Bell,$^{1,\,2,\,*}$ Chenhang~Xu,$^{1,\,2,\,*}$ Arthur~McCray,$^{3}$ Minyong~Han,$^{1,\,2}$ \\ Yicheng~Zhuang,$^{4}$ Harold~Y.~Hwang,$^{1,\,2}$ Colin~Ophus,$^{3,\,2,\,\dagger}$ and Alfred~Zong$^{1,\,2,\,5,\,\dagger}$\\[4pt]
\blfootnote{$^{*}$These authors contributed equally: H.B. and C.X.\\
$^{\dagger}$\href{mailto:cophus@stanford.edu}{cophus@stanford.edu} (C.O.) and \href{mailto:alfredz@stanford.edu}{alfredz@stanford.edu} (A.Z.)}
\footnotesize
\textit{
$^{1}$\StanfordAP\\
$^{2}$\SIMES\\
$^{3}$\StanfordMSE\\
$^{4}$\GLAM\\
$^{5}$\StanfordP}\\[4pt]
\vspace{0.25 cm}

\end{center}
\vspace{6pt}
\twocolumngrid

\section{T\lowercase{hin film surface parameterization}} \label{supp:surf_param}

In this section, the mathematical foundation of bend contour electron tomography (BCET) will be laid out in terms of thin film surface parameterization, local lattice information, and strain tensor derivation. A thin film can be represented by a 2D surface embedded in 3D space, $\mathbf{R}(u,v)$. Here, $u,v\in[0,1]$ are parametric surface coordinates. Both the height and strain information are contained in this mathematical description. To compute the local lattice vectors, we first compute the tangent vectors, $\mathbf{u}, \mathbf{v}$, given by,
\begin{equation}
\mathbf{u}(u,v) = \frac{1}{u_0}\frac{\partial \mathbf{R}}{\partial u},\hspace{1 cm} \mathbf{v}(u,v) = \frac{1}{v_0}\frac{\partial \mathbf{R}}{\partial v}.
\end{equation}
These vectors are normalized not by their local amplitude, but by global amplitudes, $u_0, v_0$, to retain the strain information. These global amplitudes are calculated based on a flat reference configuration surface. We can now define a local coordinate system captured by a $3\times3$ coordinate matrix, $S(u,v)$, given by
\begin{equation}
    S(u,v) =
\begin{bmatrix}
| & | & | \\
\mathbf{u}(u,v) & \mathbf{v}(u,v) & \hat{\mathbf{n}}(u,v) \\
| & | & |
\end{bmatrix}.
\end{equation}
Here, $\hat{\mathbf{n}}$ is a unitary surface normal vector $\hat{\mathbf{n}} \parallel (\mathbf{u} \times \mathbf{v})$. More generally, the surface normal vector need not be a unit vector if out-of-plane strain is present. Here, we focus on in-plane strain, so we are only concerned about the unit vector $\hat{\mathbf{n}}$; the case for out-of-plane strains is considered in Sec.~\ref{supp:c_lattice_recon}. The $S(u,v)$ matrix defines a coordinate system at each point on the $\mathbf{R}(u,v)$ surface with the tangent and normal vectors. 

The real space lattice vectors at each local coordinate are readily computed from the columns of a related $3\times3$ matrix $D(u,v)$, defined as
\begin{equation}
    D(u,v)=S(u,v)\,UM,
\end{equation}
where the columns of the $3\times3$ matrix $M$ are the real space lattice vectors, $\mathbf{a},\mathbf{b},\mathbf{c}$ [see Fig.~\ref{fig:forward_sim}(a)], expressed in an arbitrarily chosen reference frame of the material system. The matrix, $U$, is a global rotation matrix that transforms the lattice vectors into the experimental reference frame, and it allows for the situation where $\mathbf{a}$ is not necessarily parallel to $\mathbf{u}$. The reciprocal lattice vectors, $\mathbf{a}^*$, $\mathbf{b}^*$, $\mathbf{c}^*$ are then the columns of $2\pi D^{-1}(u,v)$.

To compute the in-plane strain of a given surface function, nonlinear strain theory is used \cite{holzapfel__nonlinear}. The undeformed configuration, $\mathbf{R}_0(u,v)$, is defined as a flat surface with $S_0(u,v) \equiv I$ for all $u,v$ coordinates, where $I$ is the identity matrix. The deformed configuration is the surface function, $\mathbf{R}(u,v)$. The linear transformation from the reference frames of the undeformed configuration to the deformed configuration is the deformation gradient tensor, $F(u,v)$, defined as
\begin{align}
 F(u,v)\,S(u,v) &= S_0(u,v) \equiv I, \text{ or}\\
 F(u,v) &= [S(u,v)]^{-1}.
\end{align}
The deformation gradient tensor encompasses the rotational, shear, and scaling transformations from the frame of the undeformed reference surface to the frame of the deformed surface. To remove the rotational components of the deformation gradient, the right Cauchy strain tensor can be found with 
\begin{align}
C(u,v) = F(u,v)^T F(u,v).     
\end{align}
The Green-Lagrange strain tensor is then computed with 
\begin{align}
 \varepsilon (u,v) = \frac12\left[C(u,v) - I\right].   
\end{align}
The in-plane portion of this tensor is used as the metric of strain in the BCET algorithm.

\section{B\lowercase{end contour as an approximate iso-gradient contour}} \label{supp:bc_math}

\begin{figure*}[htb!]
    \includegraphics[width=1\linewidth]{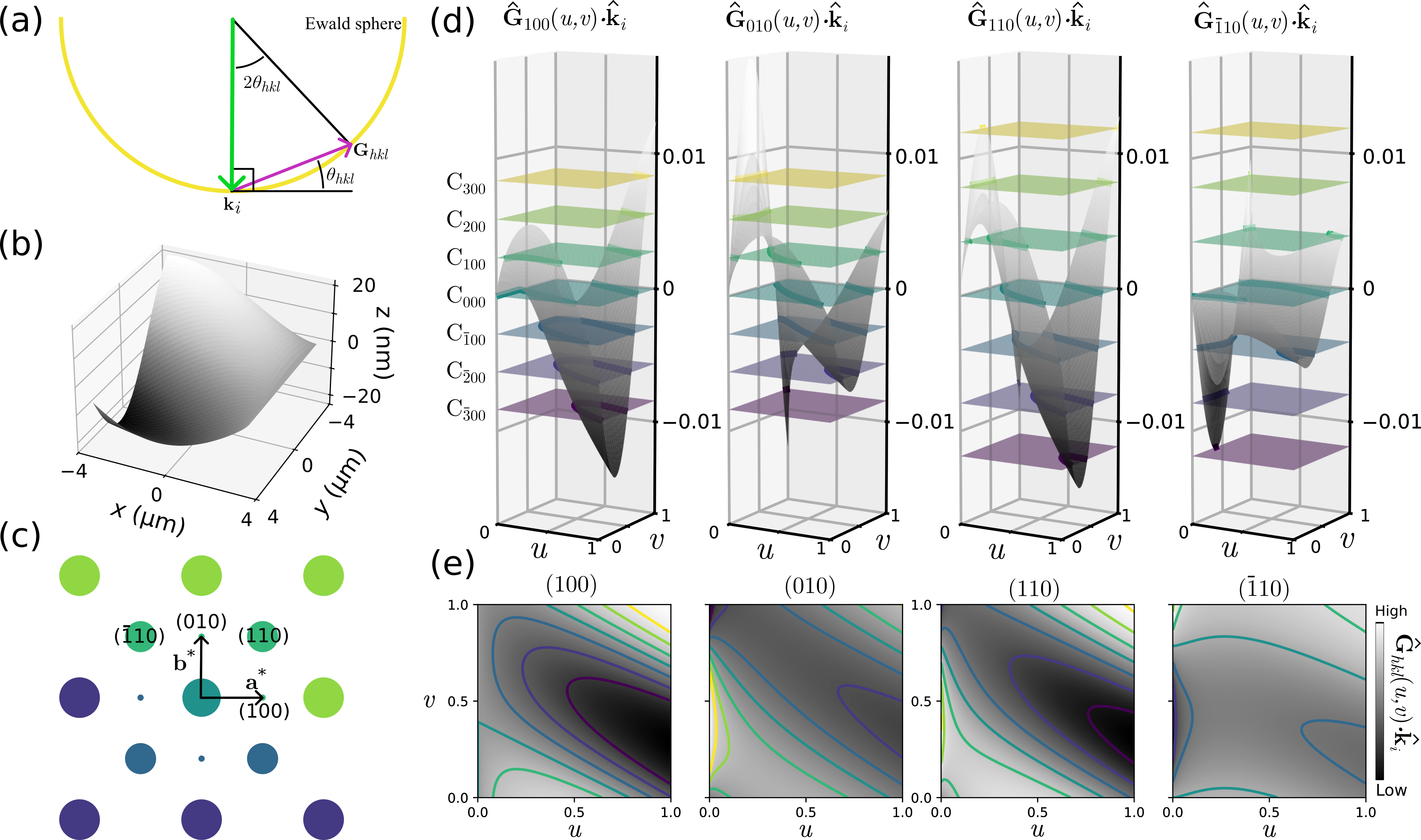}
    \caption{\textbf{Bend contour condition explained.} (a)~Schematic of the scattering geometry. A reciprocal space lattice vector $\mathbf{G}_{hkl}$ that meets the diffraction condition is tilted by the Bragg angle $\theta_{hkl}$. (b)~An example thin film surface topography with zero strain. (c)~Schematic diffraction pattern of SrTiO$_3$. The diameter of the circles is proportional to the expected intensity. (d)~Plotting-based solution of the bend contour condition in Eq.~\eqref{eq:bc_condition}. The monotonic surfaces are the generalized gradients, $\hat{\mathbf{G}}_{hkl} (u,v) \cdot \hat{\mathbf{k}}$, plotted for (100), (010), (110), and ($\overline{1}10$) Bragg orders. In color, we show $C_{hkl}$ as horizontal planes. The intersection of the horizontal planes and generalized gradient surfaces yields the solutions to the bend contour condition and traces the position of the bend contours in the absence of lattice strain. (e)~Bend contour locations on an image. The generalized gradient surfaces in (d) are depicted as images, with the surface intersections with $C_{hkl}$ shown as contours. Contour colors correspond to different Bragg orders, with the same coloring scheme in panels~(c) and (d).}
    \label{fig:bc_condition}
\end{figure*}

Bend contours can be thought of as the location on the film where the diffraction condition is met, but this misses a key insight into why this contrast forms 1D contours on a continuous crystalline surface. Here we introduce a more concrete mathematical description of bend contour contrast which can explain this intrinsic feature. For the diffraction condition to be met, the reciprocal lattice vector, $\mathbf{G}_{hkl} \equiv h \mathbf{a}^* + k \mathbf{b}^* + l \mathbf{c}^*,$
needs to be elevated from the plane perpendicular to the incident electron wavevector, $\mathbf{k}_i$, into the Ewald sphere by the Bragg angle, $\theta_{hkl}$ [Fig.~\ref{fig:bc_condition}(a)]. In a surface with curvature or strain, $\mathbf{G}_{hkl}$ is a function that depends on the position on the surface, $\mathbf{G}_{hkl}(u,v)$. As shown in Fig.~\ref{fig:bc_condition}(a), the diffraction condition of $\mathbf{G}_{hkl}(u,v)$ can be expressed as
\begin{align}
\frac{\pi}{2} - \angle (\mathbf{G}_{hkl}(u,v), \mathbf{k}_i) &= \arcsin(\hat{\mathbf{G}}_{hkl}(u,v) \cdot \hat{\mathbf{k}}_i)\notag \\
&= \theta_{hkl},    
\end{align}
where $\theta_{hkl}$ is the Bragg angle. Using Bragg's law, 
\begin{equation}
\sin[\theta_{hkl}(u,v)] = \frac{|\mathbf{G}_{hkl}(u,v)|}{2|\mathbf{k}_i|},    
\end{equation}
we get the \emph{bend contour condition},
\begin{equation}\label{eq:bc_condition}
\hat{\mathbf{G}}_{hkl}(u,v) \cdot \hat{\mathbf{k}}_i = \frac{|\mathbf{G}_{hkl}(u,v)|}{2|\mathbf{k}_i|}.
\end{equation}
At any coordinates $(u,v)$ where Eq.~\eqref{eq:bc_condition} holds, the diffraction condition is met and bend contours will be observed. Graphically, both the left and right hand sides of Eq.~\eqref{eq:bc_condition} define a 2D surface parameterized by $(u,v)$ for a given $(hkl)$ order. The intersection of two 2D surfaces therefore results in a 1D contour. If $\mathbf{R}(u,v)$ is a smooth surface, we expect both left and right hand sides of Eq.~\eqref{eq:bc_condition} to be smooth surfaces of $(u,v)$. Under this condition, the resulting bend contour is a continuous 1D curve, which either forms closed loops or terminates at the edge of the domain. 

Let us gain some intuition about the left and right hand sides of Eq.~\eqref{eq:bc_condition}. The term $\hat{\mathbf{G}}_{hkl}(u,v)$ on the left hand side is computed based on a linear transformation of the vectors defining the tangent plane at $(u,v)$. Hence, we can view $\hat{\mathbf{G}}_{hkl}(u,v) \cdot \hat{\mathbf{k}}_i$ as some generalized gradient of the surface at $(u,v)$ projected into specific directions as defined by the Bragg order $(hkl)$ and the incident electron direction $\hat{\mathbf{k}}_i$. We will henceforth call $\hat{\mathbf{G}}_{hkl}(u,v) \cdot \hat{\mathbf{k}}_i$ \emph{generalized gradient}. As for the right hand side of Eq.~\eqref{eq:bc_condition}, in the special case of zero strain, it  becomes independent of $(u,v)$, so we can define
\begin{equation}
C_{hkl} \equiv \frac{|\mathbf{G}_{hkl}(u,v)|}{2|\mathbf{k}_i|},    
\end{equation}
which are simply horizontal planes in the $(u,v)$ coordinates.

The graphical solutions to the bend contour condition in Eq.~\eqref{eq:bc_condition} in the zero-strain case are visualized through Fig.~\ref{fig:bc_condition}(b--e). For the SrTiO$_3$ film surface $\mathbf{R}(u,v)$ shown in Fig.~\ref{fig:bc_condition}(b), we compute the generalized gradients for four $(hkl)$ peaks [see Fig.~\ref{fig:bc_condition}(c)]. For the right hand side of Eq.~\eqref{eq:bc_condition}, constant height surfaces are drawn at $C_{hkl}$ values for each $(\eta h, \eta k, \eta l)$ triplet [see peak selections in Fig.~\ref{fig:bc_condition}(d)], where $\eta$ is an integer. Here, $\eta$ is needed because the $\hat{\mathbf{G}}_{hkl}(u,v)$ term in Eq.~\eqref{eq:bc_condition} is a unit vector, so all $\hat{\mathbf{G}}_{\eta h, \eta k, \eta l}(u,v)$ are identical. Finally, bend contour contrast of the $(\eta h,\eta k, \eta l)$ order is present at the intersection between the two surfaces defined by the left and right hand sides of Eq.~\eqref{eq:bc_condition}, as plotted in Fig.~\ref{fig:bc_condition}(e). In this case, bend contours can be understood as constant-height contours of the generalized gradient, i.e., iso-gradients.

The left hand side of Eq.~\eqref{eq:bc_condition} for the bend contour condition also points out two important invariants. Because the left hand side only depends on the relative angle between the incident electron direction $\hat{\mathbf{k}}_i$ and the reciprocal lattice unit vector $\hat{\mathbf{G}}_{hkl}(u,v)$, the bend contour position will be unchanged under a global rotation of the crystal if the rotation axis is $\hat{\mathbf{k}}_i$ or $\hat{\mathbf{G}}_{hkl}(u,v)$. More specifically, if the crystal rotates around $\hat{\mathbf{k}}_i$, bend contours of all orders remain unchanged in their locations. If the crystal rotates around $\hat{\mathbf{G}}_{hkl}(u,v)$, all $(\eta h, \eta k, \eta l)$ bend contours remain unchanged. These properties are useful when we construct the direct solver, as explained in more detail in Sec.~\ref{supp:general_direct_constraints}.

Moving beyond the trivial case of zero strain, we can also consider the special case of a spatially homogeneous in-plane strain. In this case, the generalized gradient on the left hand side of Eq.~\eqref{eq:bc_condition} remains unchanged, while the right side will be scaled linearly by the strain magnitude depending on the strain direction relative to the local $\mathbf{G}_{hkl}(u,v)$ direction. The isocontour planes in Fig.~\ref{fig:bc_condition}(d) will remain flat and horizontal, but their vertical spacing will contract (or expand) with tensile (or compressive) strain. In the most general case where the in-plane strain is spatially inhomogeneous, the right hand side of Eq.~\eqref{eq:bc_condition} is no longer a constant function of $(u,v)$. The bend contours no longer occur at a constant height in Fig.~\ref{fig:bc_condition}(d), but can still be defined as the intersection of two surfaces. 

Tilting the sample, which is commonly done in experiment, will add a $(u,v)$-independent constant to $\hat{\mathbf{G}}_{hkl}(u,v) \cdot \hat{\mathbf{k}}_i$, where the constant depends on the tilt axis and angle. The right hand side of Eq.~\eqref{eq:bc_condition} will be unchanged. Rotation of the sample will therefore vertically displace the curved surfaces in Fig.~\ref{fig:bc_condition}(d), changing their intersection with the planes, which will smoothly move and change the shapes of the contours. 

Fundamentally, bend contours can be formulated as the intersection of two surfaces. With small strain fields, they can be thought of as constant height contours of the generalized gradient surface, $\hat{\mathbf{G}}_{hkl}(u,v) \cdot \hat{\mathbf{k}}_i$.

\section{F\lowercase{orward model details}} \label{supp:forward_model}

\begin{figure}[t!]
    \includegraphics[width=0.95\linewidth]{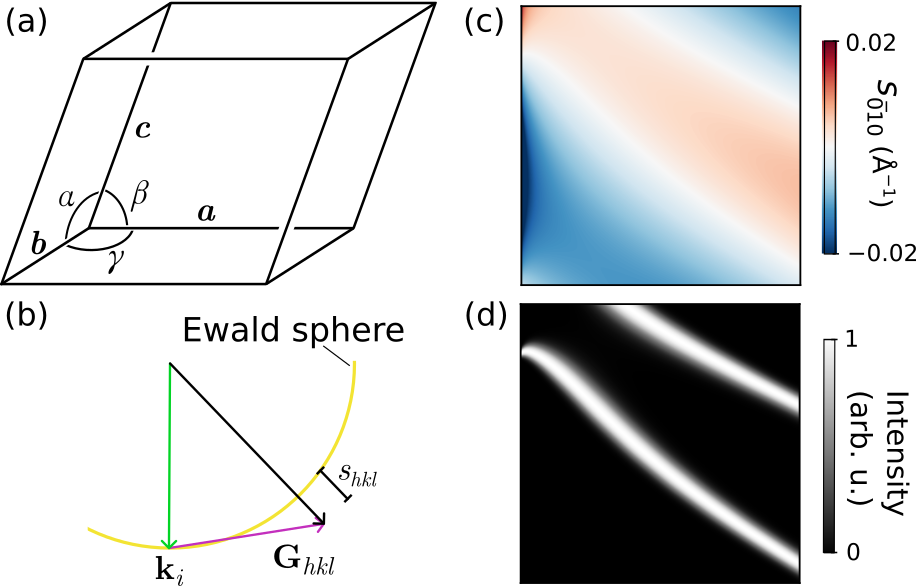}
    \caption{\textbf{Excitation error in the forward model  and intensity simulation.} (a)~The most general unit cell as defined by $a$, $b$, $c$ vector magnitudes and $\alpha$, $\beta$, $\gamma$ angles between the crystal lattice vectors. (b)~Schematic of the excitation error computed for a general $\mathbf{G}_{hkl}$ vector as the signed distance from the Ewald sphere. (c)~Excitation error, $s_{0\overline{1}0}$ over the surface $(u, v)$ coordinates. Where the excitation error crosses zero, bend contour intensity is seen in $I^{\text{df}}_{0\overline{1}0}$ in (d).}
    \label{fig:forward_sim}
\end{figure}

In this section, we describe the computational procedure by which the bend contour images are generated, namely, the forward model. We compute the TEM diffraction intensity by considering the excitation error, $s_{hkl}$, which is the minimal momentum-space distance from $\mathbf{G}_{hkl}(u,v)$ to the Ewald sphere surface \cite{williams_Transmission_2009}. We compute $s_{hkl}(u,v)$ as
\begin{equation}
s_{hkl}(u,v) = |\mathbf{k}_i| - |\mathbf{k}_i + \mathbf{G}_{hkl}(u,v)|.
\label{supp:eq:excitation_error}
\end{equation}
Due to the finite volume of the reciprocal peak and the finite thickness of the Ewald sphere, bend contours must have a finite width. We model these considerations by computing the dark-field bend contour intensity $I^{\text{df}}_{hkl}(u,v)$ using a Gaussian function with width $s_0$, 
\begin{equation}\label{eq:I_hkl_df}
    I^{\text{df}}_{hkl}(u,v) = \exp\left(-\frac{s_{hkl}^2(u,v)}{s_0^2}\right).
\end{equation}
We chose a Gaussian over a more theoretically motivated sinc function \cite{williams_Transmission_2009} because Gaussian is numerically efficient to evaluate and does not significantly change the simulated bend contour images in our thickness range. An example excitation error map and dark-field intensity map are shown in Fig.~\ref{fig:forward_sim}(c,d). At the locations where the excitation error reaches $0$, the dark-field intensity evaluates to $1$. 

\begin{figure*}[t!]
    \includegraphics[width=1\linewidth]{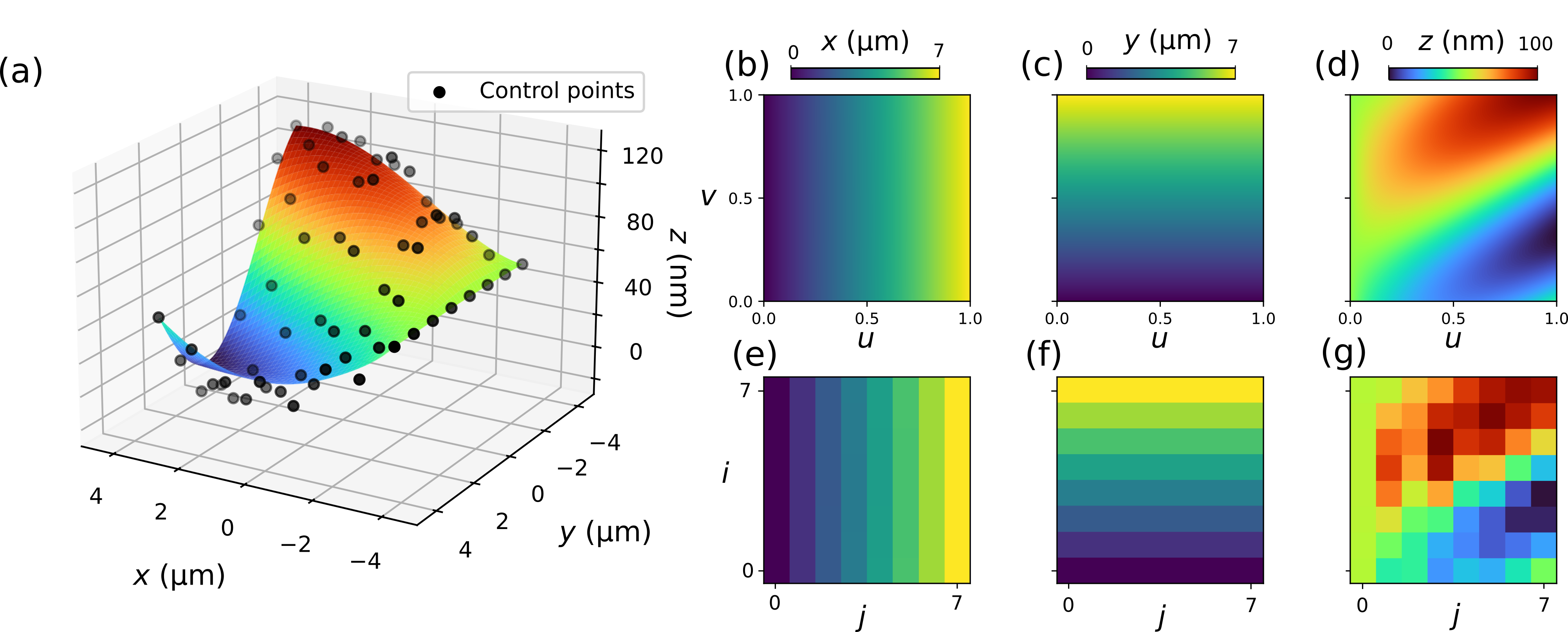}
    \caption{\textbf{B\'ezier control points and the B\'ezier surface.} (a)~Three dimensional visualization of the surface created by the control points plotted in red. (b--d)~Cartesian components of the surface plotted as images. (e--g)~Control point matrices $\mathbf{k}_{ij}$ for each Cartesian component.}
    \label{fig:figS3}
\end{figure*}

We next compute the bright-field intensity map,
\begin{align} \label{eq:I_bf_uv}
    I^{\text{bf}}(u,v) = I_0 \left[1-\sum_{h,k,l} A_{hkl}|F_{hkl}|^2I^{\text{df}}_{hkl}(u,v)\right] + I_\text{bg}.
\end{align}

Here, $h,k,l$ are all possible Miller indices up to a cutoff depending on the tilt of the sample; in addition, the case of $h = k = l = 0$ is excluded in the sum. $I_0$ is a global adjustment to match the experimental electron counts and $I_\text{bg}$ is a constant background offset due to factors such as the detector dark current. $F_{hkl}$ is the complex structure factor based on the kinematic diffraction theory while $A_{hkl}$ is an order-dependent normalization factor that also accounts for effects such as multiple scattering.

It is worth emphasizing that the quantitative outputs of BCET, including the surface topography and in-plane strain tensor, are not encoded in the intensity of bend contours but in their \emph{positions}, namely, the loci in $(u,v)$ where the excitation error vanishes [$s_{hkl}(u,v) = 0$, see Eq.~\eqref{supp:eq:excitation_error}]. The intensity-related quantities in the forward model, such as the Gaussian width $s_0$ [Eq.~\eqref{eq:I_hkl_df}], the order-dependent amplitude $A_{hkl}$ [Eq.~\eqref{eq:I_bf_uv}], and the global scale $I_0$ [Eq.~\eqref{eq:I_bf_uv}], act only to match the appearance of the contours and are treated as free nuisance parameters that are fit alongside other reconstruction parameters. Effects that redistribute or rescale diffracted intensity, including dynamical and multiple scatterings, thickness inhomogeneity, and partial beam coherence do not bias the reconstructed morphology and strain because they do not displace the contour positions even though they can modify contour contrast and width.

\section{B\lowercase{\'ezier surface parameterization}} \label{supp:Bezier}

\begin{figure*}[t!]
    \includegraphics[width=1\linewidth]{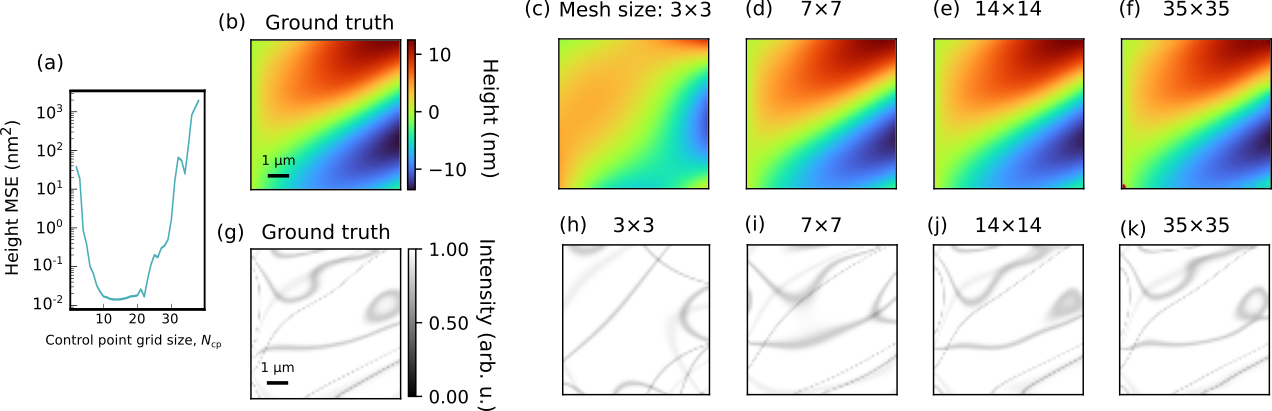}
    \caption{\textbf{Direct solver reconstruction on a simulated data set.} (a)~Height MSE for inverse solver solutions using a variable number of control points for the B\'ezier solution. On the $x$ axis, the control point number refers to $N_{\text{cp}}$, where the control point meshes are $N_{\text{cp}} \times N_{\text{cp}}$. (b)~Ground truth height map. (c--f)~Solved height maps for different numbers of control points. (g)~Ground truth bright-field images at $-5^\circ$ tilt. (h--k)~Bright-field images reconstructed using different numbers of control points.}
    \label{fig:test_direct}
\end{figure*}

For iterative optimization, using the pixel basis of the surface, which would involve optimizing each point in $\mathbf{R}(u,v)$, becomes intractable in terms of memory use and computational time for high-resolution TEM images. Instead, a different surface parameterization is needed. We found inspiration from computer graphics to use B\'ezier surfaces as our surface description \cite{piegl_Bspline_1997}. In this representation, control points $\mathbf{k}_{i,j}$ define a 2D mesh in 3D space. The surface function can be found with 
\begin{equation}
    \mathbf{R}(u,v) = \sum_{i = 0}^{N - 1} \sum_{j = 0}^{M - 1} B^n_i(u)B^m_j(v)\mathbf{k}_{i,j},
\end{equation}
which interpolates a smooth surface through the control points with arbitrary $u$-, $v$-resolution. The functions $B_{i,j}$ are Bernstein polynomials defined as,
\begin{equation}
    B_i^n(u) = \binom{n}{i}u^i(1-u)^{n-i}.
\end{equation}

For our application, B\'ezier surfaces are useful for a few reasons stemming from the observation that transformation from control points to surface coordinates is linear. For example, a matrix $Z$ can be created that linearly multiplies the control point coordinates to retrieve the surface coordinates,
\begin{equation}\label{eq:bez_matrix}
  \mathbf{R}_m = \sum_n Z_{mn} \mathbf{k}_n.
\end{equation}
Here, the indices $u,v$ of $\mathbf{R}(u,v)$ are flattened to a single index $m$ for computational convenience. Similarly, the control point indices $(i,j)$ are flattened into an index $n$. $Z_{mn}$ is computed with
\begin{equation}
  Z_{mn} \equiv Z(u,v)_{ij} = B^n_i(u)B^m_j(v).  
\end{equation}
This linear relationship enables the pyramidal optimization strategy. After convergence is reached for one mesh size during pyramidal optimization, we need to find a higher resolution control point mesh that produces the same surface as the lower resolution control point mesh. To find an optimally matched surface with $N^*$ control points instead of $N$ control points, we can equate the surface coordinates, 
\begin{equation}
    \mathbf{R}^N_m = \sum_n Z_{mn}^{N^*} \mathbf{k}_n^{N^*}.    
\end{equation}
This is a linear constraint that can be efficiently solved with least squares. As explained in Sec.~\ref{supp:general_direct_constraints}, we will also use this linear transformation property in the direct solver, where this transformation matrix multiplies the constraint matrix, enabling us to solve for the control point mesh rather than the surface pixel coordinates.

\section{D\lowercase{irect solver details}} \label{supp:general_direct_constraints}

In this section, we will explain the constraints used in the direct solver for a fully general unit cell. This constraint has to be carefully constructed to be linear in the surface height pixels, $h_m = h_{p,q}$, so we can solve it with least squares. Here, ($p,q$) are the discrete integral indices for the continuous parameterization coordinates ($u,v$), and as before, the flattened indices of $(p,q)$ is denoted by $m$. Here, $h$ is used to describe the surface height because without strain, it is a simpler parameterization of the surface than $\mathbf{R}(u,v)$.

Generally, we want to formulate constraints that can be described in the form of
\begin{align}
\mathbf{z} \cdot \mathbf{h} = \xi,     
\end{align}
where $\mathbf{h}$ is the flattened height vector over the 2D surface, $\mathbf{z}$ is some vector, and $\xi$ is a scalar, which together describe a general linear constraint on $\mathbf{h}$. This constraint formulation from bend contour contrast is nontrivial to devise because the reciprocal lattice vectors are a nonlinear function of the surface height. However, the surface normal unit vector, 
\begin{align}\label{eq:n_hat_parallel_condition}
\hat{\mathbf{n}} \parallel (\mathbf{a} \times \mathbf{b}) \parallel \mathbf{c}^*, 
\end{align}
can be linearized in the limit of small ripple amplitude. Therefore, we formulate the diffraction condition as a constraint on the normal vector. 

Let us first review the discussion in the main text, where in the absence of strain and in the special case of a cubic lattice system, a convenient constraint to use for the observation of a bend contour at $\mathbf{G}_{hk0}$ is, 
\begin{align}\label{eq:cubic_constraint}
  \mathbf{G}_{hk0} \cdot \hat{\mathbf{n}} = 0,  
\end{align} 
where we know that $\mathbf{G}_{hk0}$ intersects with the Ewald sphere; by assuming one global in-plane rotation, $\mathbf{G}_{hk0}$ can be uniquely determined. We can solve for the normal vector from the height map $h_{p,q}$ with the cross product between two non-parallel vectors in the tangent plane to find,
\begin{equation}\label{eq:normal_vector}
    \mathbf{n} = - h_x\hat{x} - h_y\hat{y} + \hat{z},
\end{equation}
where $\hat{x}$, $\hat{y}$, $\hat{z}$ are the unit vectors in the Cartesian coordinate directions and
\begin{align}
    h_x &= \frac{(h_{p+1,q} - h_{p-1,q})}{2\Delta x},\\
    h_y &= \frac{(h_{p,q + 1} - h_{p, q-1})}{2\Delta y}.    
\end{align}
Here, $\Delta x$ and $\Delta y$ are the scalings from $p,q$ to real space Cartesian coordinates on the grid. The constraint from Eq.~\eqref{eq:cubic_constraint} in the matrix form reads
\begin{equation}\label{eq:G_h_0}
    \mathbf{G}_{hkl} \cdot \begin{pmatrix}
        (h_{p-1,q} - h_{p+1,q})/(2\Delta x) \\
        (h_{p,q-1} - h_{p,q+1})/(2\Delta y) \\
        1
    \end{pmatrix} = 0.
\end{equation}

For general space groups, Eq.~\eqref{eq:G_h_0} does not always hold true. The general constraint can be formulated as a fixed angle between $\mathbf{G}_{hk0}$ and $\hat{\mathbf{n}}$, where the angle is not necessarily $\pi/2$. This constraint can be understood from the conditions for diffraction and bend contour formation:  when the crystal is rotated around $\mathbf{G}_{hkl}$, the diffraction condition for that order is unchanged. To formulate the general equation for this constraint based on the height map, we must find the constant angle, denoted by $\psi_{hkl}$, between $\mathbf{G}_{hkl}$ and $\mathbf{\hat{n}}$. Let us first define the three angles, $\alpha^*$, $\beta^*$ and $\gamma^*$ as, 
\begin{align}
  \alpha^* &= \angle (\mathbf{b}^*,\mathbf{c}^*),\\
  \beta^* &= \angle(\mathbf{a}^*, \mathbf{c}^*),\\
  \gamma^* &= \angle (\mathbf{a}^*, \mathbf{b}^*).  
\end{align}
To find $\psi_{hkl}$, we compute the dot product between $\mathbf{\hat{n}}$ and $\mathbf{G}_{hkl}$, yielding
\begin{equation}
  \cos(\psi_{hkl}) = \frac{h a^* c^* \cos{\beta^*} + k b^* c^* \cos{\alpha^*} + l (c^*)^2  }{|\mathbf{G}_{hkl}|c^*},  
\end{equation}
where we utilized the relation in Eq.~\eqref{eq:n_hat_parallel_condition}. Here $a^*\equiv|\mathbf{a}^*|$, $b^*\equiv|\mathbf{b}^*|$, and $c^*\equiv|\mathbf{c}^*|$. As a sanity check, if $l = 0$ and $\alpha = \beta = \pi/2$, we retrieve the expected result of $\cos(\psi_{hk0}) = 0$. 

\begin{figure}[b!]
    \includegraphics[width=1\linewidth]{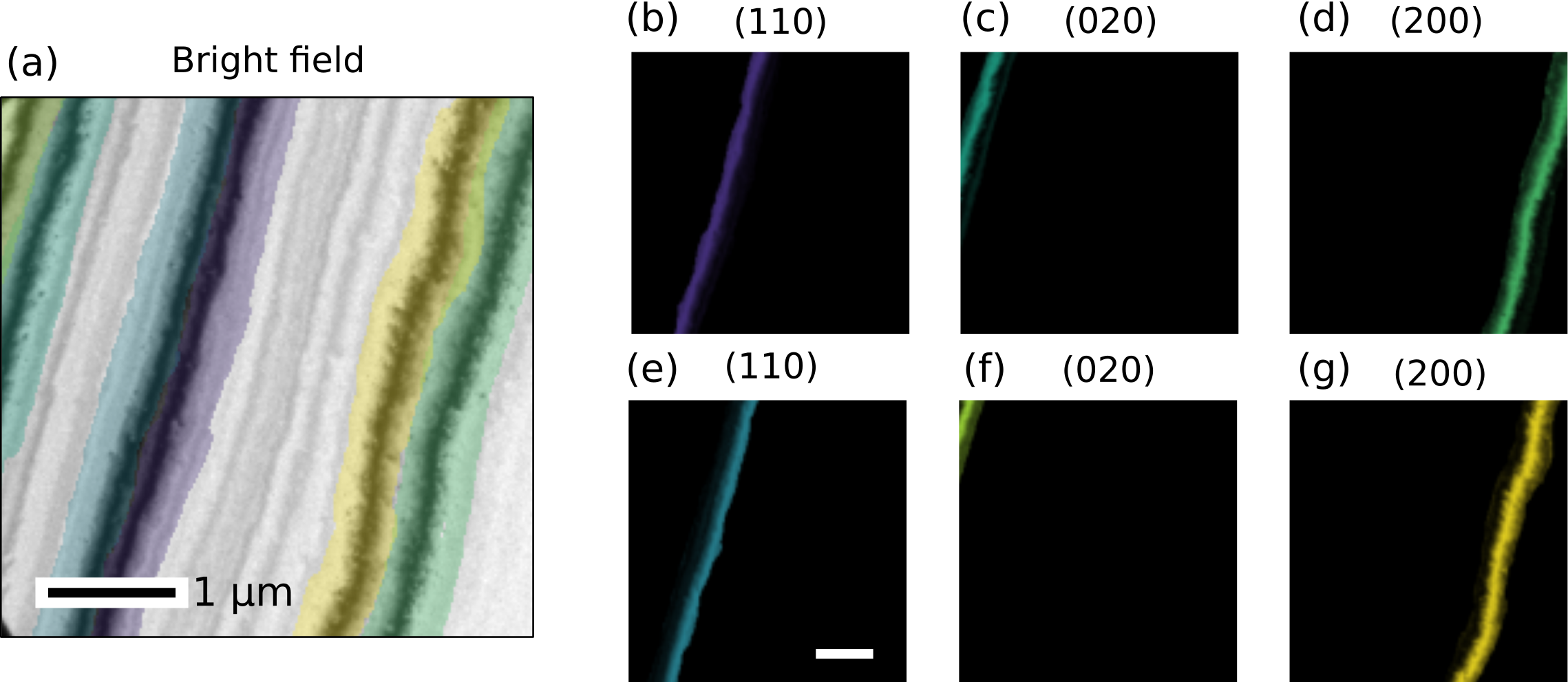}
    \caption{\textbf{Bright-field image converted to virtual dark-field images.} (a)~Example 300~keV bright-field image of a 30-nm-thick SrTiO$_3$ film on a 2,000 mesh copper grid. Manual image labeling was used to generate masks and virtual dark-field images (b--g)~of the strong bend contours.}
    \label{fig:dark_field_labelling}
\end{figure}

\begin{figure}[t!]
    \includegraphics[width=0.8\linewidth]{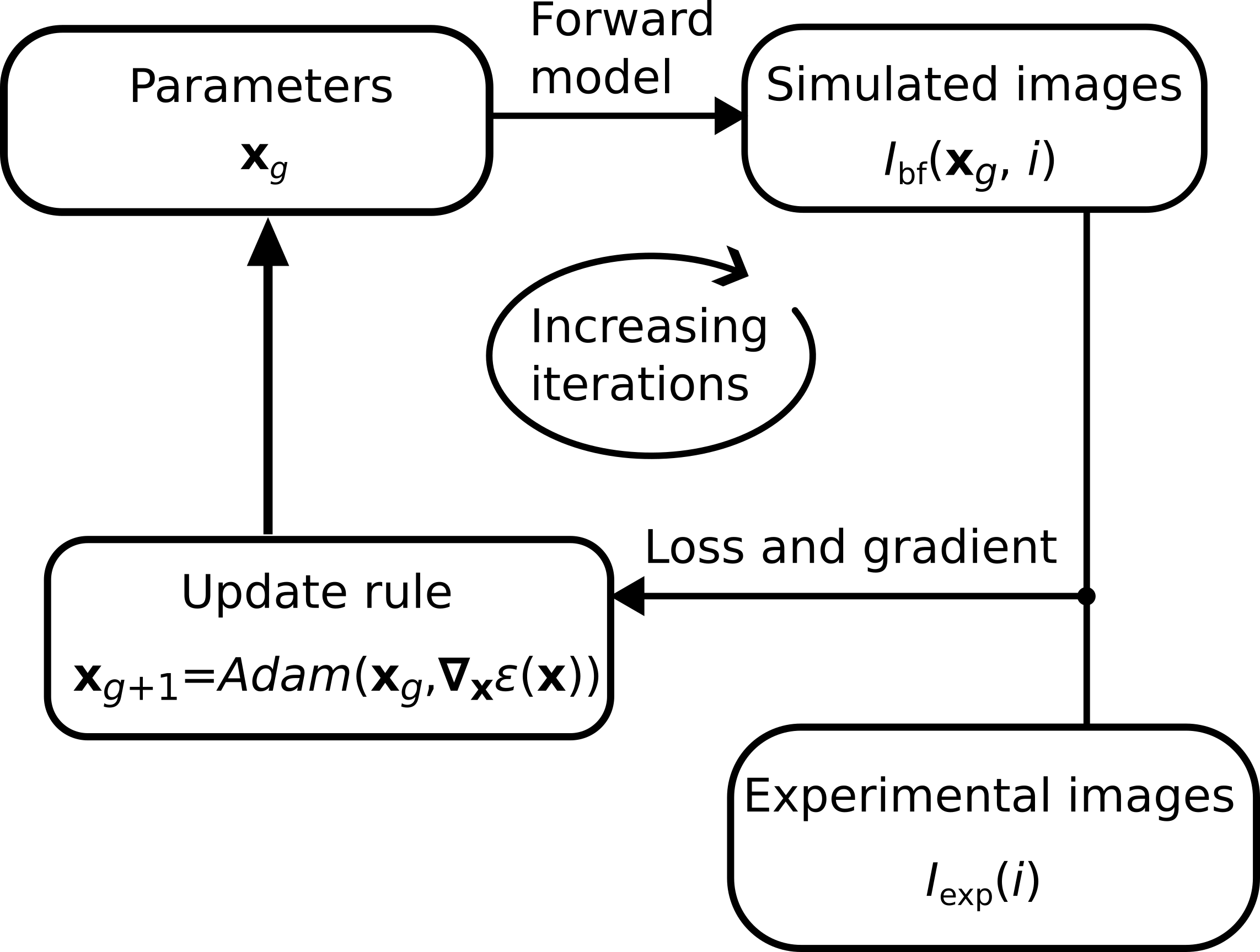}
    \caption{\textbf{Optimization loop.} Parameters that define the sample surface and experimental conditions at the $g$th iteration ($\mathbf{x}_g$) are improved with an iteration loop. First, the forward model is used to compute simulated bright-field images [$I_{\text{bf}}(\mathbf{x}_g, i)$]. These are compared with the experimental images to form the MSE loss. The gradient of the loss is used to update the parameters with the \textit{Adam} update rule.}
    \label{fig:optimization_loop}
\end{figure}

We are now ready to formulate the general constraint on the height map, $h_{p,q}$. We first find the \emph{unit} normal vector,
\begin{equation}
    \hat{\mathbf{n}} = \frac{- h_x\hat{x} - h_y\hat{y} + \hat{z}}{\left(h_x^2 + h_y^2 + 1\right)^{1/2}} \approx - h_x\hat{x} - h_y\hat{y} + \hat{z}. 
\end{equation}
In the limit where $h_x^2, h_y^2 \ll 1$, we find that $\mathbf{\hat{n}}$ is approximately equal to $\mathbf{n}$ in Eq.~\eqref{eq:normal_vector}. This is true when the slope of the wrinkles in the thin film is much less than 1. Although higher-order terms could be retained for larger slopes, the small-slope approximation ensures linearity, making the reconstruction tractable via linear least-squares. The general constraint then becomes,
\begin{equation} \label{eq:direct_method_constraint}
    \mathbf{G}_{hkl} \cdot \begin{pmatrix}
        (h_{p-1,q} - h_{p+1,q})/(2\Delta x) \\
        (h_{p,q-1} - h_{p,q+1})/(2\Delta y) \\
        1
    \end{pmatrix} = \cos{(\psi_{hkl})}.
\end{equation}

Coming back to our original goal to constrain the surface, for each dark-field pixel, for a certain order $(h,k,l)$, we find the $\mathbf{G}_{hkl}$ vector that intersects with the Ewald sphere, and has the assumed in-plane rotation. Then, we rewrite Eq.~\eqref{eq:direct_method_constraint} in the form of $\mathbf{z}\cdot\mathbf{h} = \xi$ for efficient linear computation, where $\mathbf{z}$ and $\xi$ can be directly derived from $\mathbf{G}_{hkl}$. These constraints are stacked into a matrix $X$ and as a whole can be written as 
\begin{align}\label{eq:linear_constraint_on_h}
X\mathbf{h} = \boldsymbol\xi.   
\end{align} 
Here, the dimension of $X$ is the number of constraints by the number of pixels in $h_{p,q}$, and $\boldsymbol\xi$ is a vector of stacked $\xi$'s with length equal to the number of constraints.

As mentioned in Sec.~\ref{supp:Bezier}, we do not need to solve all the constraints in the pixel basis, $h_{p,q}$. Instead, we can solve it in the B\'ezier basis, which provides a few key advantages. Using the $z$ coordinate of Eq.~\eqref{eq:bez_matrix}, we can rewrite $\mathbf{h}$ as
\begin{equation}
    \mathbf{h} = Z \mathbf{k}^z,    
\end{equation}
where $\mathbf{k}^z$ is the $z$ coordinate of the control point mesh. The direct solver constraint then becomes,
\begin{equation} \label{eq:direct_method_constraint_bez}
    (XZ)\mathbf{k}^z = \boldsymbol{\xi}.
\end{equation}

\begin{figure*}[t!]
    \includegraphics[width=0.9\linewidth]{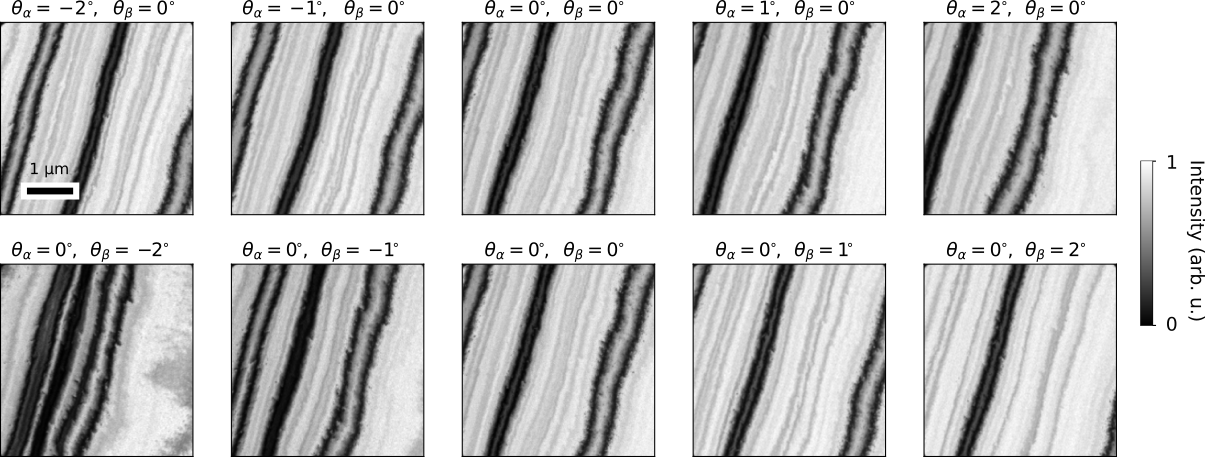}
    \caption{\textbf{Experimental SrTiO$_\text{3}$ tilt series.} Ten selected experimental bright-field images at different $\theta_\alpha$ and $\theta_\beta$ tilts. Bend contours translate and distort across the film with tilt.}
    \label{fig:experimental_images_example}
\end{figure*}

Solving Eq.~\eqref{eq:direct_method_constraint_bez} has the added computational advantage over solving Eq.~\eqref{eq:linear_constraint_on_h} because the number of control points is usually orders of magnitude fewer than the number of pixels. Using the B\'ezier solution basis also has the implicit benefit of providing regularization because B\'ezier surfaces are naturally smooth.

To validate the direct solver, we performed a simulated reconstruction on an example surface without strain. The ground truth height of a known analytic surface function and the corresponding bright-field image from the forward model are shown in Fig.~\ref{fig:test_direct}(b,g). We enforced the direct solver constraints on the hot pixels of the thresholded simulated dark-field images, and solved the resulting matrix in the B\'ezier basis. The resulting height and bright-field images are shown in Fig.~\ref{fig:test_direct}(c--f) and (h--k). We computed the mean squared error (MSE) of the height for each control point mesh and report the result in Fig.~\ref{fig:test_direct}(a). We observe an underfitting range from 2 $\times$ 2 to roughly 10 $\times$ 10 control points, followed by a wide appropriate fitting range from 10 $\times$ 10 to 25 $\times$ 25 control points, and an overfitting regime above 25 $\times$ 25 control points.

\begin{figure}[b!]
    \includegraphics[width=0.99\linewidth]{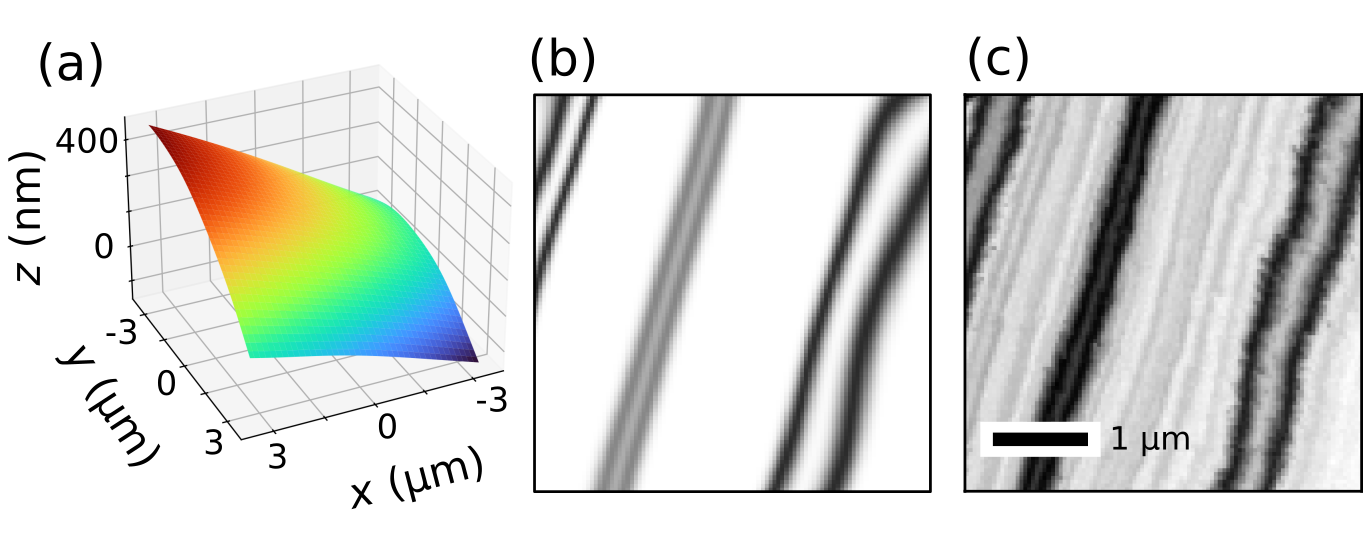}
    \caption{\textbf{Results of the direct solver on a measured SrTiO$_\text{3}$ film.} (a)~Reconstructed height map from the direct solver with a $10 \times 10$ control point grid. (b)~Simulated bright-field image from the direct solver result. (c)~Experimental bright-field image at the same tilt. The peak width and peak intensity were selected manually to match the experimental images.}
    \label{fig:fig3_direct_method}
\end{figure}

\begin{figure*}[tb!]
    \includegraphics[width=0.9\linewidth]{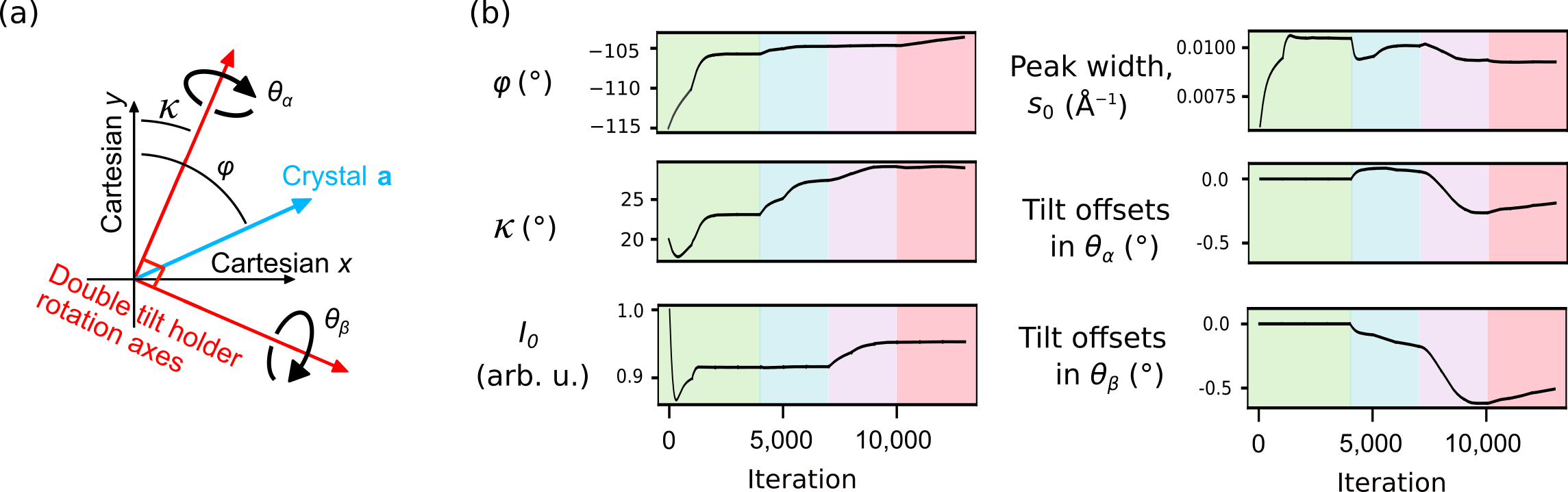}
    \caption{\textbf{Details of the iterative reconstruction on SrTiO$_\text{3}$.} (a)~Schematic of geometric parameterization of quantities that are iteratively optimized. (b)~A select set of the optimized parameter values throughout the optimization for the experimental SrTiO$_3$ reconstruction. Tilt offsets in $\theta_\alpha$ and $\theta_\beta$ are optimized for each bright-field image in the tilt series, but only those for one image are plotted for visual clarity. The tilt offsets are fixed until the second phase (blue shade).}
    \label{fig:fig3_extra}
\end{figure*}

\section{D\lowercase{ark field labeling}}\label{supp:dark_field_labelling}

For the input to the direct solver, we must know the order of each bend contour. A straightforward way to achieve this task is to perform dark-field mapping of all peaks at each tilt. While possible, in practice it becomes unrealistic to do this. Instead, while acquiring a bright-field tilt series, we performed dark-field mapping only when unknown bend contours appear upon tilting. In this modality, some post processing is required. In particular, manual image labeling is performed on each tilt series image to assign each contour to an $(hkl)$ triplet based on the slow continuous deformation of the bend contours during sample rotation and comparing known bend contour orders from a previous tilt. In Fig.~\ref{fig:dark_field_labelling}, an example manual labeling is shown for the SrTiO$_3$ sample area, which is reconstructed in this work. We labeled six contour orders from the bright-field images, resulting in masks which we used to create virtual dark-field images. The masking itself is facilitated by standard interactive segmentation tools. In our workflow, we used an open-source data-labeling tool called \textit{Label Studio} \cite{Label_Studio}. This tool provides brush/polygon selection and propagation of labels between adjacent frames, which substantially reduce the per-image effort relative to labeling from scratch.

Two types of errors may occur during manual labeling. First, the order of a bend contour may be mislabeled, but such errors are easy to verify by checking the consistency between labels in nearby tilts and are thus not a primary concern. Second, the exact shape of the mask may not be consistent across different label trials. However, because the direct-solver constraints are derived from a skeletonized, single-pixel-wide representation of each virtual dark field image rather than the exact mask boundary (see Sec.~\ref{exp_recon_details}), the solver is relatively insensitive to the precise shape of the manual mask. Furthermore, this manual labeling step only informs the direct solver, and therefore only informs the initial conditions of the BCET iterative reconstruction. While this labeling is an important and required step, the final converged solution is not sensitive to small variations of the initial conditions.

We recognize that the labeling step is the principal manual bottleneck of the direct solver, but automation is possible given rapid advances of automated and autonomous experimentation in (S)TEM, with demonstrations of self-driving microscopes that autonomously navigate image space, correct aberrations, and select acquisition targets under varying conditions \cite{roccapriore_Autonomous_2022, kalinin_Machine_2022}. In our context, the two automatable pieces are (i)~acquisition of the dark-field tilt series used for order cross-referencing, which fits naturally within existing self-driving TEM frameworks, and (ii)~contour masking and order assignment, for which the continuous frame-to-frame deformation of contours makes propagation-based tracking and learned segmentation well suited.

\section{I\lowercase{terative optimization loop}} \label{supp:optimization_loop}

For the iterative method, a computational loop is evaluated at each iteration to minimize the loss, $\epsilon(\mathbf{x})$ in Eq.~\eqref{equation:cost_function}. This loop is shown in Fig.~\ref{fig:optimization_loop}. First, the parameters $\mathbf{x}_g$, which consists of the surface control points and geometric/experimental parameters where $g$ is the iteration count, are sent through the forward model to compute the bright-field images $I_\text{bf}(\mathbf{x}_g, i)$. The index $i$ denotes the different tilt angles. Simulated images are then compared with the experimental images with MSE to form the loss function. We tried multiple image similarity metrics, such as cross correlation, but we chose MSE based on a good balance between simplicity and performance. The loss gradient is efficiently computed with a backpropagation step through the computational graph, which is implemented with automatic differentiation in \textit{PyTorch} \cite{paszke_pytorch_2019a}. The gradients are used to descend the loss landscape with the \textit{Adam} update rule \cite{kingma_adam_2015}. After a fixed number of iterations, we terminate the optimization loop. The loss and parameter stability is then evaluated to see if the solution is converged. Hyperparameters such as \textit{Adam} learning rates were manually adjusted based on multiple simulated and experimental reconstructions, balancing speed and successful convergence.

\section{E\lowercase{xperimental reconstruction details}}\label{exp_recon_details}

In our experiment, we employed an FEI Titan TEM with 300~keV acceleration voltage at a magnification of 1750$\times$. We employed a Gatan K3 direct electron detector to capture 4K resolution bright-field images. We collected 1~s exposure bright-field tilt series images of a freestanding 30-nm-thick SrTiO$_3$ film on a 2,000 mesh copper TEM grid (see Sec.~\ref{supp:sample_prep} for sample preparation details). We used a double-tilt holder with angles $\theta_\alpha$, $\theta_\beta$ where the tilt axes are perpendicular to each other. See Fig.~\ref{fig:fig3_extra}(a) for the experimental geometry and Fig.~\ref{fig:experimental_images_example} for example bright-field images. We performed the tilt series by fixing one of the angles and rotating the other, with $0.5^\circ$ steps from $-5^\circ$ to $5^\circ$ in each direction. During the tilt series, we employed dark-field mapping if new bend contours appeared, and performed manual dark-field labeling as described in Sec.~\ref{supp:dark_field_labelling}. 

For the direct solver, the virtual dark-field images were sent through a skeletonize image filter \cite{walt_Scikitimage_2014} to extract the single pixel-thick central bend contour line as a binary image. Each nonzero pixel is converted to a constraint and solved with the direct solver (see Sec.~\ref{supp:general_direct_constraints}). We solve for the B\'ezier surface control points $\mathbf{k}_{i,j}$ in the Cartesian basis, $\hat{x}, \hat{y}, \hat{z}$, where $\hat{z} \parallel -\mathbf{k}_i$, and $\hat{x}$, $\hat{y}$ are parallel to the image axes.  For the input to the direct solver, the relationship between the image pixel axes, the rotation axes, and the image scale must be approximately known, as well as the global in-plane rotation of the reciprocal space vectors. In our case, the $(110)$ order peaks do not move significantly with $\theta_\beta$, even up to rotations of $15^\circ$, allowing us to ascertain that the $\theta_\beta$ rotation axis is approximately parallel to the $\mathbf{G}_{110}$ vector. Additionally, we use the known shape of the cylindrical hole in the TEM grid, as well as the projected intensity through the hole at the largest accessible tilt angle ($\theta_\alpha = \pm40^\circ$) to approximately determine the $\theta_\alpha$ axis. The image scale is determined by the known diameter of the 2,000-mesh TEM grid. Using the constraints from the virtual dark-field images, and the described sample/tilt holder geometry, we use the direct solver with a $10 \times 10$ control point grid to solve for the surface topography in the absence of strain. The resulting surface and example bright-field image are shown in Fig.~\ref{fig:fig3_direct_method}(a,b). Comparing the direct solver result and the corresponding experimental image [Fig.~\ref{fig:fig3_direct_method}(b,c)], the direct solver indeed yields a reasonable guess as an input for the iterative solver.

For the iterative solver, the surface solution from the direct solver in Fig.~\ref{fig:fig3_direct_method}(a) is used as a starting point. In practice, the circular field of view is cropped to an inner square, which streamlines the convergence for the iterative solver. Additionally, parameters like the peak width ($s_0$) and global intensity ($I_0$) are manually selected to match the bright-field image, which act as the initial guess for the iterative solver. These parameters are then refined with the iterative method. 

\begin{figure*}[t!]
    \includegraphics[width=0.9\linewidth]{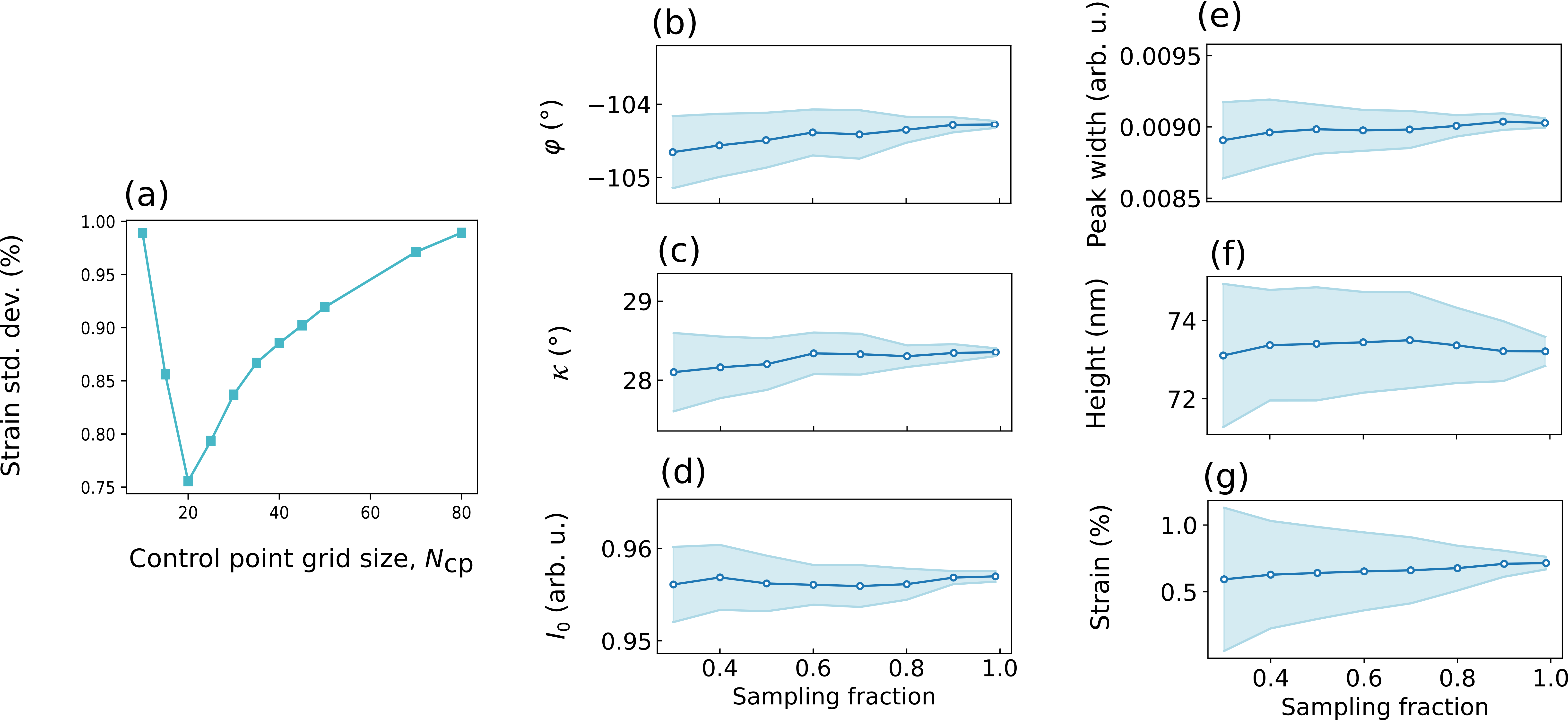}
    \caption{\textbf{Cross validation results for the experimental SrTiO$_\text{3}$ reconstruction.} (a)~Control point-dependent leave-half-out cross validation for the strain reconstruction. Each data point represents the standard deviation in strain (area averaged) over 40 individual reconstructions with varying control point grids of size $N_{\text{cp}} \times N_{\text{cp}}$. (b--g)~Cross validation standard deviation compared for varying sampling fractions for a select set of the reconstruction variables. The circular markers are the mean value over reconstructions, while the shades represent the standard deviation over reconstructions. For spatially-resolved variables, like height and strain, the circular markers are the mean absolute deviation from the mean value, where both means are taken over the spatial dimensions. For strain, an additional mean is taken over all strain tensor elements. The plot in (g) is a different way to visualize the same data presented in Fig.~\ref{fig:3}(k).}
    \label{fig:cross_correlation_fig}
\end{figure*}

In particular, for the multiphase optimization shown in Fig.~\ref{fig:fig3_extra}(b), in the first region (green), the peak intensity adjustment $A_{hkl}$ of the dominant bend contours is optimized, and so is the Gaussian width ($s_0$), in-plane crystal axis rotation ($\varphi$), and the $\theta_\alpha$ rotation axis ($\kappa$). The $\theta_\beta$ axis is fixed to be perpendicular to the $\theta_\alpha$ axis [see the experimental geometry schematic in Fig.~\ref{fig:fig3_extra}(a)]. In this first region, the control point mesh number $N_\text{cp}$ is slowly scaled up from $10$ to $20$ using the pyramidal method, as decided from the leave-half-out cross validation procedure described in Sec.~\ref{supp:precision}, where the cross validation is minimized at 20 control points [Fig.~\ref{fig:cross_correlation_fig}(a)].

In the second (blue) region of Fig.~\ref{fig:fig3_extra}(b), tilt offsets in $\theta_{\alpha,\beta}$ are relaxed, which are included because of inherent error in the reported angle of the rotation stages in the TEM. For example, we observed that in two bend contour images taken at the same reported angles tens of minutes apart, the bend contours are translated with respect to each other, indicating a drift in the reported tilt angle. The optimized values of tilt offset stay within roughly $\pm0.6^\circ$ without any constraint.

In the third (pink) region of the multiphase procedure in Fig.~\ref{fig:fig3_extra}(b), 50 more peaks are added from $h, k\in[-7,7]$ and $l = 0$. Due to computational speed considerations, only the peaks that contribute intensity to the bright-field images are included. The initial intensity adjustment $A_{hkl}$ of these peaks is chosen to be 0, constrained to be the same value for symmetrical $(h,k,l)$, and refined through iterative optimization. In the final region, the strain is relaxed, meaning that the control points in the B\'ezier parametrization are allowed to move laterally, in $x$ and $y$. 

\section{D\lowercase{etermination of the optimal number of} B\lowercase{\'ezier control points for the iterative reconstruction}}\label{supp:control_pt_num_determine}

\begin{figure*}[t!]
    \includegraphics[width=0.85\linewidth]{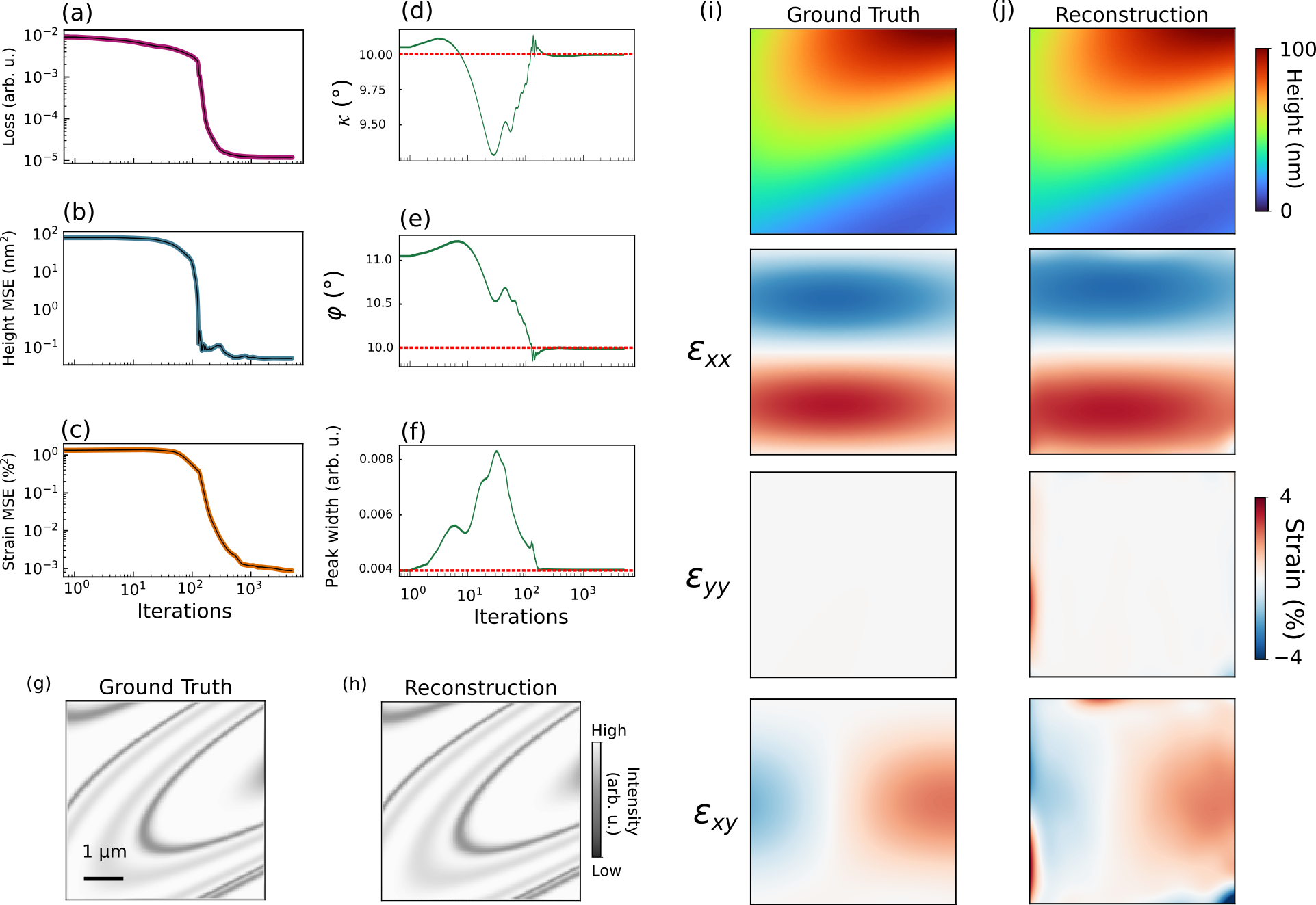}
    \caption{\textbf{Iterative reconstruction evaluated on a simulated dataset.} (a--c)~The loss, height MSE and strain MSE through iterative optimization. (d--f)~The in-plane angle of the sample rotation axis, crystal vectors, and peak width during the optimization. The dotted red line is the ground truth value. (g,h)~Ground truth and reconstructed bright-field images at zero tilt. (i)~Ground truth height and strain tensor maps. (j)~Reconstructed height and strain maps.}
    \label{fig:reconstruct_accuracy}
\end{figure*}

In this section, we use cross validation \cite{kohavi_Study_1995, arlot_Survey_2010} to decide the optimal number of control points. This approach generally relies on repeated random sub-sampling of the experimental images, performing independent reconstructions from different data subsets, and comparing the resulting solutions. 

Before applying cross-validation to determine the optimal number of control points, it is important to consider the bias–variance tradeoff associated with model complexity. Increasing the number of control points improves the ability of the model to resolve fine spatial features, but also increases sensitivity to noise and the risk of overfitting, resulting in higher variance. Conversely, using too few control points limits the expressive capacity of the model, leading to underfitting and increased bias. This tradeoff implies that an intermediate number of control points exists that balances model flexibility and stability, providing an optimal representation of the surface.

To find this control point mesh size, we performed a leave-half-out cross validation strategy. In particular, starting from the same initial $\mathbf{x}$, we split the available data into two halves that are randomly sampled, and we performed a reconstruction on each half separately. This procedure is repeated multiple times, and the final strain distributions are compared between reconstructions. High variance in the reconstructed strains using different data would suggest either under- or over-fitting to the data, while minimum variance suggests appropriate fitting. By repeating this procedure for a different number of control points, the optimal control point number can be surmised. In this case, because we are confident in the height map results, we start our cross validation from the converged result of the optimization without strain. 

We performed leave-half-out cross validation with control point mesh size from $10\times 10$ to $80 \times 80$, with 40 reconstructions at each control points number. The cross validation metric, chosen to be the standard deviation of the strain over reconstructions, averaged over the surface, is plotted in Fig.~\ref{fig:cross_correlation_fig}(a). The plot shows a standard cross-validation behavior: with a low number of control points, we are not able to accurately capture the experimental data, and we are therefore underfitting, resulting in higher strain variance. If we use too many control points, the surface will overfit to the data subset and we will again see high strain variance. We find a minimal cross validation standard deviation of $0.76\%$ at $N_\text{cp} = 20$ (i.e., 20$\times$20 grid). Therefore, we use this control point number in the final reconstruction.

\section{A\lowercase{ssessing the reconstruction precision}}\label{supp:precision}

In order to evaluate the precision of our reconstructed solution in the absence of the ground truth, we performed a related cross validation sub-sampling analysis as in Sec.~\ref{supp:control_pt_num_determine} but instead of varying control point number, we use a variable sampling fraction. Namely, we choose a sub-sample of the tilt series images with sampling fraction from 0.3 to 0.975. At each sampling fraction, we perform 40 reconstructions with independent datasets. The variability among these reconstructions at a given random sampling fraction provides a quantitative basis for selecting the optimal regularization parameter and evaluating the statistical precision of the reconstruction.

In Fig.~\ref{fig:cross_correlation_fig}(b--g), we provide the scaling of the cross validation error with variable sampling fraction and a fixed control point number for a select collection of optimized parameters. To assess the precision of the reconstructed parameters in Fig.~\ref{fig:cross_correlation_fig}(b--g), we evaluate the cross-validation standard deviation at the largest sampling fraction, which provides the closest estimate of the statistical precision of the final reconstruction. For parameters with physically meaningful absolute magnitudes like strain or height deviation, the cross-validation standard deviation can be compared directly to the characteristic parameter value to assess statistical significance. For example, for the strain, the cross validation standard deviation falls below the mean deviation magnitude at a sampling fraction of $0.5$ [see the shaded region and circular markers in Fig.~\ref{fig:cross_correlation_fig}(g)], indicating a statistically meaningful result for our reconstruction. For the height in Fig.~\ref{fig:cross_correlation_fig}(f), the cross validation standard deviation is on the order of 1~nm, which is well below the mean height near 73~nm. For parameters with arbitrary reference offsets, such as global orientation angles, precision can only be quantified in absolute terms. In our case, angular parameter cross validation standard deviation values are less than $0.2^\circ$, indicating a stable reconstruction that is well constrained by the data.

\begin{figure}[htb!]
    \centering
    \includegraphics[width=0.95\linewidth]{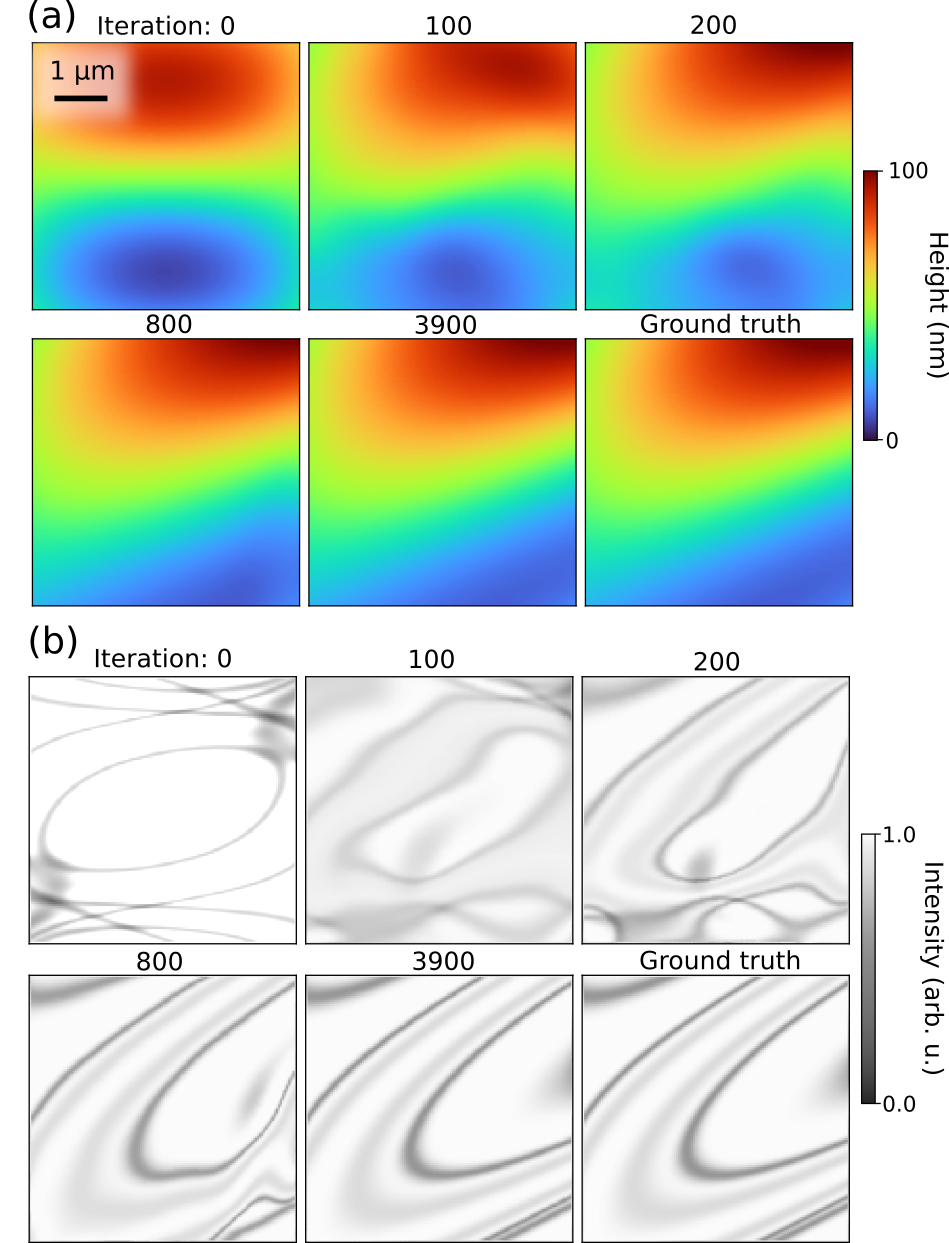}
    \caption{\textbf{Surface height and bend contours during iterative reconstruction.} (a)~Height maps at different iteration counts compared with the ground truth height. (b)~Bright-field images at zero tilt for different iteration counts. These images can be compared with the ground truth bright-field image at zero tilt.}
    \label{fig:test_data_results_pt2}
\end{figure}

\section{A\lowercase{ssessing the reconstruction accuracy}}\label{supp:accuracy}

To assess the accuracy of the BCET method, we performed a validation using simulated data from a known ground truth surface. Specifically, we defined an analytical surface geometry, simulated the corresponding bend contour bright-field images, and applied the BCET reconstruction algorithm to recover the surface and strain from these images. The reconstructed surface was then directly compared to the ground truth to quantify reconstruction accuracy. The ground truth surface was defined by the analytical function below, with characteristic length scale $d = $ \SI{7}{\micro\metre},
\begin{align}
X(u,v) &= d \left(u'(u) - 1/2 \right), \\
Y(u,v) &= d\left(v'(v) - 1/2 \right),\\
Z(u,v) &= A \left(\sqrt{u' + B} - \sqrt{B} \right) \notag\\
       &~~~~~\cdot\sin(2u' + 2v')/6 + Cu',
\end{align}
where $A, B, C$ are chosen constants and
\begin{align}
    u'(u) &= u + D\sin{(\pi u)}^2\text{e}^{-(v-v_0)^2/\lambda},\notag\\
    v' &= v.
\end{align}
The $u'(u)$ term produces the tensile/compressive strain centered at $v_0$. The bright-field images from this surface were simulated at a 10$^\circ$  clockwise rotation angle from the $y$ axis, with 20 tilt series images from $-5^\circ$ to $5^\circ$ in a single tilt axis. A selected image at zero tilt is shown in Fig.~\ref{fig:reconstruct_accuracy}(g). To test the limits of the inverse solver, we use a starting guess surface that is generated from a combination of analytical functions with zero strain,
\begin{align}
  X_0(u,v) &= d(u - 1/2),\label{eq:sim_surface_guess_X}\\
  Y_0(u,v) &= d(v - 1/2),\label{eq:sim_surface_guess_Y}\\
  Z_0(u,v) &= A_1 \sin\left[\pi X (u,v)+ d/2\right] \notag\\  &~~~~~~~~\cdot \sin\left[2\pi(Y(u,v) + d/2)\right].\label{eq:sim_surface_guess_Z}
\end{align}
This starting guess is significantly further from the ground truth than what the direct solver provides. To start the optimization, the functions in Eqs.~\eqref{eq:sim_surface_guess_X}--\eqref{eq:sim_surface_guess_Z} were fit with control points using Eq.~\eqref{eq:bez_matrix} to define the initial control point mesh. The initial geometric parameters such as rotation, width, in-plane rotation were set slightly off their ground truth value. We sent this initial guess through the iterative optimization loop 5,000 times, resulting in the reconstruction in Fig.~\ref{fig:reconstruct_accuracy}.

Unlike experimental reconstruction, we now have access to the height MSE and strain MSE [Fig.~\ref{fig:reconstruct_accuracy}(b,c)], which reach extremely small values. In both simulated and experimental cases we have access to the loss, which is shown in this case in Fig.~\ref{fig:reconstruct_accuracy}(a). The parameters such as peak width, tilt axis, and in-plane lattice angles over optimization can be seen in [Fig.~\ref{fig:reconstruct_accuracy}(d--f)], which converge to their ground truth values. The surface height, strain and bright-field images closely match the ground truth [Fig.~\ref{fig:reconstruct_accuracy}(g--j)]. 

In Fig.~\ref{fig:test_data_results_pt2}, we display the surface height maps [Fig.~\ref{fig:test_data_results_pt2}(a)] and bright-field images for different iteration counts [Fig.~\ref{fig:test_data_results_pt2}(b)]. The B\'ezier surfaces quickly approach the ground truth with reasonable agreement after only 100 or 200 iterations, but their bright-field images differ significantly, showing the large sensitivity of bend contour contrast to the topography/strain.

Next, interested in the scaling behavior of the BCET algorithm with experimental noise, we augment all bright-field pixels with Poisson noise due to electron counting statistics. In particular, we augment each pixel value by applying Poisson counting statistics,
\begin{align}
    I = P(N_e I_o).
\end{align}
Here, $I_o$ is the normalized original pixel value so that it falls in the range of $[0,1]$, $P(\Omega)$ denotes a Poisson random variable with mean $\Omega$, $N_e$ is the electron counts per pixel, and $I$ is the resulting pixel value. By performing the same simulated reconstruction as described in this section but for variable $N_e$, we demonstrate the robustness of BCET against realistic experimental noise conditions; see Fig.~\ref{fig:4} and the accompanying discussion in the main text.

\section{A\lowercase{ssessing the reconstruction uniqueness}}
\label{supp:uniqueness}

Any inverse-problem approach to structural reconstruction raises the question of whether its solution is unique: whether the experimental data is sufficient to select a single surface and strain field, or whether many distinct configurations reproduce the same bend-contour images equally well. The local tilt and in-plane strain enter the bend-contour condition [Eq.~\eqref{eq:bc_condition}] in a coupled fashion, so that a single bright-field image underconstrains the problem. The degeneracy is broken by acquiring images across a range of tilts, but the minimum amount of tilt data required to reach a unique solution is not obvious \textit{a~priori}. In this section, we develop an empirical test for assessing reconstruction uniqueness and use it to quantify how the reconstruction fidelity depends on the number of tilt images and the total tilt range.

\subsection{An empirical identifiability criterion}

\begin{figure}[b!]
    \centering
    \includegraphics[width=0.75\columnwidth]{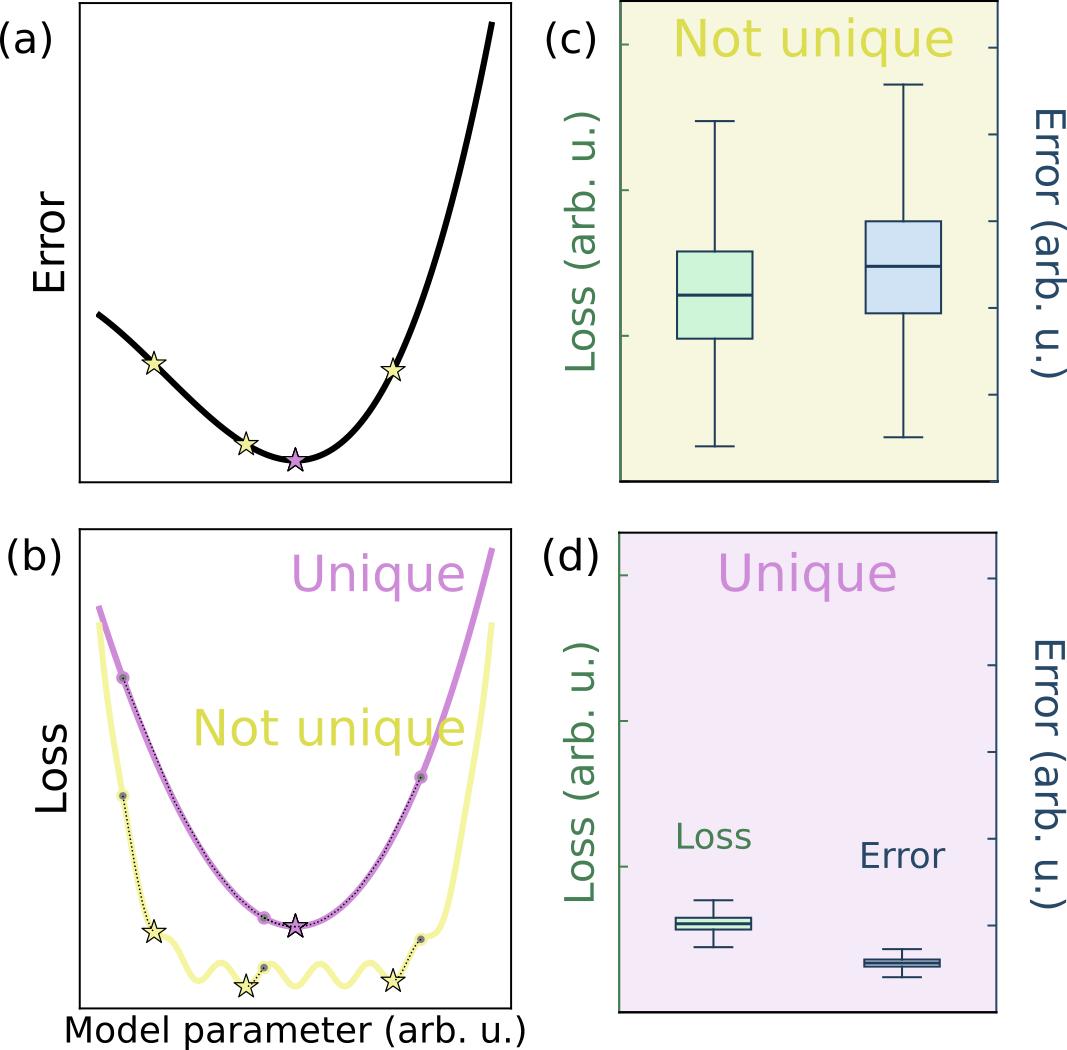}
    \caption{\textbf{Schematic illustration of the identifiability criterion used to assess reconstruction uniqueness.} (a)~A schematic of the ground-truth error (distance from the known solution) as a function of a model parameter, showing a unique global minimum (purple star). (b)~Schematic loss landscapes for two cases: a well-posed, unique reconstruction (purple curve) with a single sharp minimum, and a non-unique reconstruction (yellow curve) with a flat loss landscape or a loss landscape containing multiple minima, where random starting points (dots) converge to different final parameter values (stars). (c)~Schematic loss and ground-truth error across many random restarts for the non-unique case are visualized as box plots: both ground-truth error and loss are broadly distributed, with high variance across restarts. Note that a low loss does not guarantee a low error in this case. (d)~Corresponding statistics for the unique case: both loss and error are tightly concentrated, and low loss reliably indicates low error, confirming that the global minimum has been found.}
    \label{fig:recon_uniqueness_cartoon}
\end{figure}

\begin{figure*}[htb!]
    \centering\includegraphics[width=0.68\linewidth]{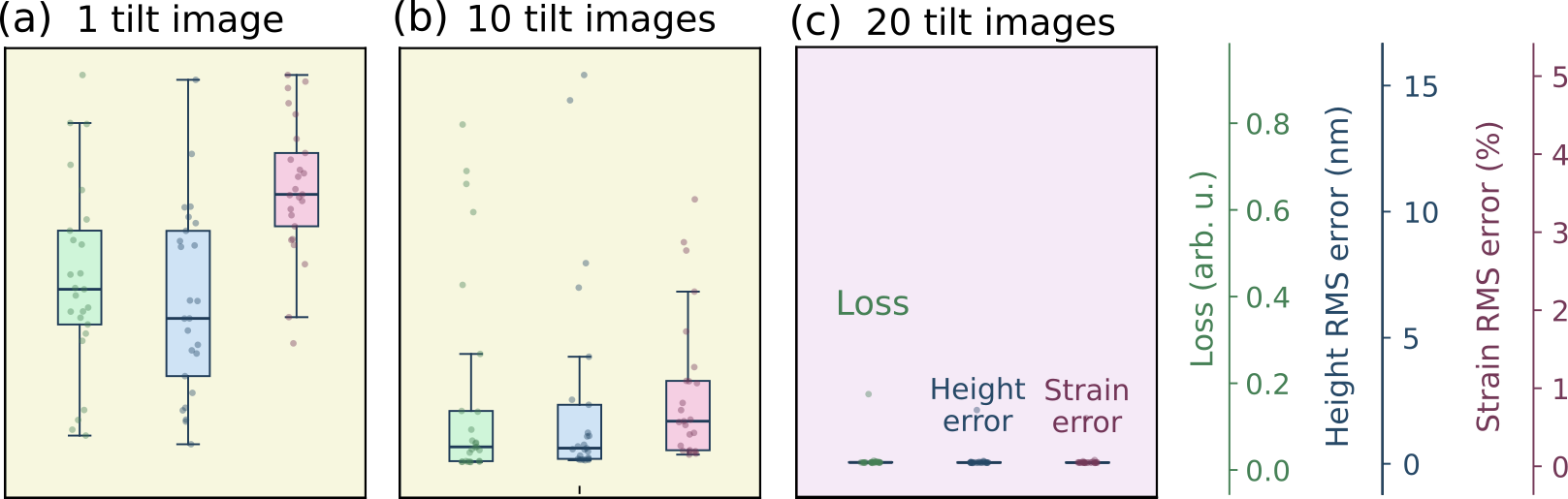}
    \caption{\textbf{Empirical identifiability test as a function of tilt image count.} (a--c)~Final loss (green), height RMS error (blue), and strain RMS error (pink) from 25 independent reconstructions, each initialized from a different randomly sampled starting point, shown as box plots for (a)~1, (b)~10, and (c)~20 tilt images used to define the loss function. Background shading indicates the inferred uniqueness regime: yellow denotes non-unique reconstructions [broad distributions across restarts, analogous to panel~(c) of Fig.~\ref{fig:recon_uniqueness_cartoon}] and purple denotes unique reconstructions [tightly concentrated distributions, analogous to panel~(d) of Fig.~\ref{fig:recon_uniqueness_cartoon}]. With only 1 tilt image, all three metrics are widely distributed, indicating that the loss landscape contains multiple solutions consistent with the data. By 20 tilt images, both the loss and the ground truth errors collapse to narrow distributions across all 25 restarts, demonstrating that the reconstruction is well-posed and converges to a unique solution. Box plots visualize the interquartile range (IQR) within the box, the median is shown as the central line in the box, whiskers extended with caps extend to 1.5 times the IQR, and points show the individual trials.}\label{fig:recon_uniqueness_data}
\end{figure*}

Because our forward model is nonlinear and the loss landscape is non-convex, we assess uniqueness empirically rather than analytically. The key idea, sketched in Fig.~\ref{fig:recon_uniqueness_cartoon}, is that a unique, well-posed reconstruction corresponds to a loss landscape with a single dominant minimum, so that independent optimizations started from different random initial conditions all converge to the same solution. A non-unique reconstruction, by contrast, possesses multiple loss minima of comparable depths, and random restarts scatter across distinct solutions with widely varying ground-truth errors [Fig.~\ref{fig:recon_uniqueness_cartoon}(a,b)]. This distinction produces a directly observable signature: in the non-unique case, the final loss and the ground-truth error are both broadly distributed across restarts, and a low final loss does not guarantee a small error because the optimizer may have settled in a spurious minimum [Fig.~\ref{fig:recon_uniqueness_cartoon}(c)]. In the unique case, both quantities are tightly concentrated, and a low loss reliably implies a
low error [Fig.~\ref{fig:recon_uniqueness_cartoon}(d)]. Crucially, this criterion can be evaluated without knowing the true solution and without constructing a specific competing configuration: the spread of outcomes across random restarts alone diagnoses the transition from the non-unique to the unique regime. When a ground-truth surface is available, as in the simulated tests below, we additionally verify that the concentrated-loss regime coincides with a small ground-truth error.

\subsection{Dependence on the number of tilt images}
\begin{figure*}[htb!]
    \centering
    \includegraphics[width=0.64\linewidth]{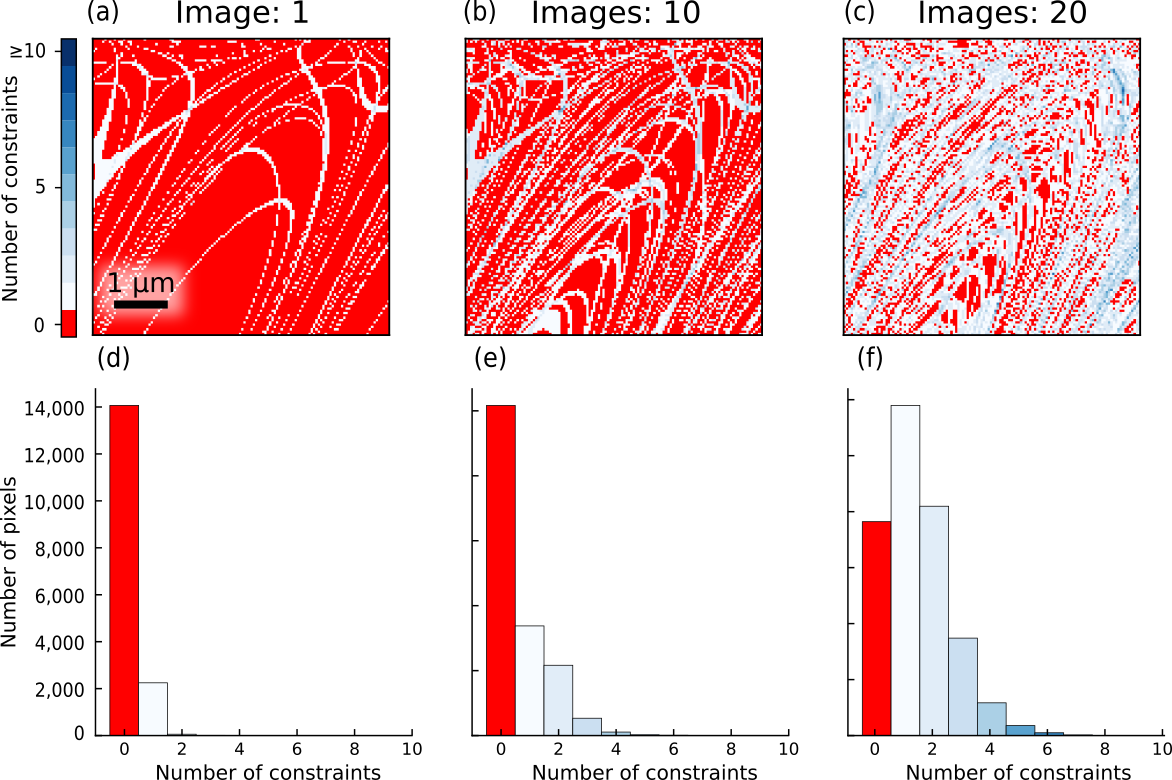}
    \caption{\textbf{Spatial coverage of bend contour constraints as a function of tilt image count.} (a--c)~Spatially-resolved maps of the number of bend contours that pass through each pixel, accumulated over tilt series of (a)~1, (b)~10, and (c)~20 images. The color scale indicates the number of constraints per pixel, with red marking pixels traversed by no bend contour (zero constraints) and progressively darker blue indicating more constraints. As the number of tilt images increases, more of the field of view is covered by at least one contour, and pixels accumulate constraints from multiple nonparallel orders. (d--f)~Histograms of the number of constraints per pixel corresponding to the maps in (a--c), respectively. With a single image~(d), most pixels receive zero constraints (red bar), and few receive more than one, leaving the surface largely underdetermined. By 20 images~(f), the zero-constraint population is substantially reduced and the distribution shifts toward multiple constraints per pixel, providing the redundant, multi-angle coverage required for a well-posed reconstruction. Some small regions are still unconstrained in (f), but by utilizing spatial regularization, the correct height/strain can still be reconstructed in these regions.}
    \label{fig:recon_constraints_n_tilts}
\end{figure*}

We first applied this criterion to determine how many tilt images are needed for a unique reconstruction for a fixed tilt range. Using the simulated ground-truth dataset of Sec.~\ref{supp:accuracy}, we performed 25 independent reconstructions from randomly sampled starting points for tilt series of 1, 10, and 20 images spanning the same $\pm 5^\circ$ angular range, and recorded the final loss, height RMS error, and strain RMS error for each restart. The resulting distributions are shown as box plots in Fig.~\ref{fig:recon_uniqueness_data}. With a single tilt image [Fig.~\ref{fig:recon_uniqueness_data}(a)], all three quantities are broadly distributed across the 25 restarts, the signature of a non-unique loss landscape in which many distinct height--strain combinations are compatible with the limited data. At 10 images [Fig.~\ref{fig:recon_uniqueness_data}(b)], the distributions narrow appreciably but retain noticeable variance. By 20 images [Fig.~\ref{fig:recon_uniqueness_data}(c)], the loss, height error, and strain error all collapse to tight distributions across every restart, indicating that the reconstruction has entered the unique regime: the data alone select a single solution, independent of initialization. Our experimental SrTiO$_3$ dataset of 40 images therefore lies well within this unique regime, providing a substantial margin of overdetermination relative to the transition.

This behavior can be understood intuitively by counting how many bend-contour constraints fall on each pixel in the field of view as the tilt series grows. Figure~\ref{fig:recon_constraints_n_tilts}(a--c) shows spatially-resolved maps of the number of bend contours passing through each pixel, accumulated over 1, 10, and 20 tilt images, with the corresponding histograms in Fig.~\ref{fig:recon_constraints_n_tilts}(d--f). With a single image, the majority of pixels are traversed by no bend contour at all, leaving the surface largely underdetermined [Fig.~\ref{fig:recon_constraints_n_tilts}(a,d)]. As the number of images increases, the bend contours sweep across the field of view, the zero-constraint population shrinks, and pixels progressively accumulate constraints from multiple nonparallel orders [Fig.~\ref{fig:recon_constraints_n_tilts}(c,f)]. At 20 tilt images, the majority of the surface is constrained, with small regions having zero constraints. These regions can still be reliably reconstructed with surface smoothness regularization given by the B\'ezier parameterization.

\subsection{Dependence on the tilt range}

\begin{figure*}[htb!]
    \centering
    \includegraphics[width=0.7\linewidth]{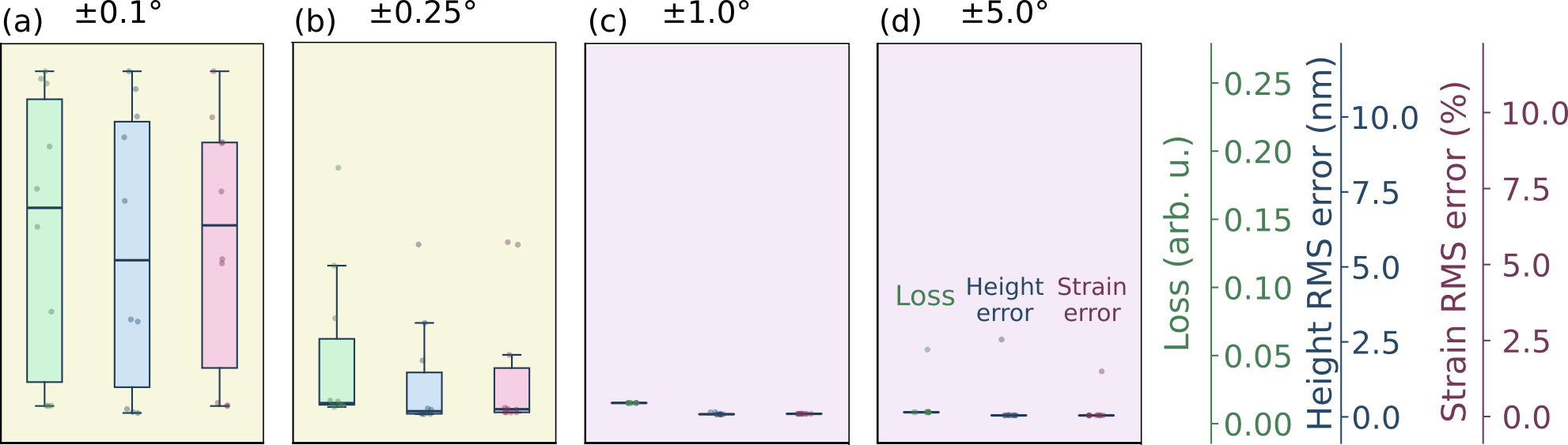}
    \caption{\textbf{Empirical identifiability test as a function of the tilt range.} Final loss (green), height RMS error (blue), and strain RMS error (pink) from 25 independent reconstructions initialized from random starting points, shown as box plots for tilt ranges of (a)~$\pm 0.1^\circ$, (b)~$\pm 0.25^\circ$, (c)~$\pm 1.0^\circ$, and (d)~$\pm 5.0^\circ$, each with 20 tilt images distributed uniformly across the range. Background shading follows the uniqueness convention of Fig.~\ref{fig:recon_uniqueness_cartoon}: yellow indicates a non-unique reconstruction regime with broad distributions across restarts, and purple indicates a unique regime with tightly concentrated distributions. All three metrics collapse to narrow distributions by $\pm 1.0^\circ$.}
    \label{fig:supp:tilt_range_uniqueness_data}
\end{figure*}

 The total angular range spanned by the tilt series also determines identifiability. To isolate its effect, we repeated the restart analysis at a fixed image count, performing 25 independent reconstructions for tilt ranges of $\pm 0.1^\circ$, $\pm 0.25^\circ$, $\pm 1.0^\circ$, and $\pm 5.0^\circ$, each with 20 images distributed uniformly across the range. To mimic a double-tilt holder, 10 images were acquired by varying $\theta_\alpha$ at fixed $\theta_\beta$, and 10 by varying $\theta_\beta$ at fixed $\theta_\alpha$. The resulting distributions of loss, height RMS error, and strain RMS error are shown in Fig.~\ref{fig:supp:tilt_range_uniqueness_data}. At the smallest ranges, $\pm 0.1^\circ$ and $\pm 0.25^\circ$, all three metrics are broadly distributed across restarts, indicating that the limited angular variation provides insufficient constraint to distinguish between distinct surface and strain configurations. By $\pm 1.0^\circ$, the distributions collapse to narrow bands, and they remain tightly concentrated at $\pm 5.0^\circ$, showing that the reconstruction has entered the unique regime. The full $\pm 5^\circ$ range with 40 images used in our experiment thus provides a comfortable margin above this threshold.
 
\begin{figure}[t!]
    \centering
    \includegraphics[width=1\linewidth]{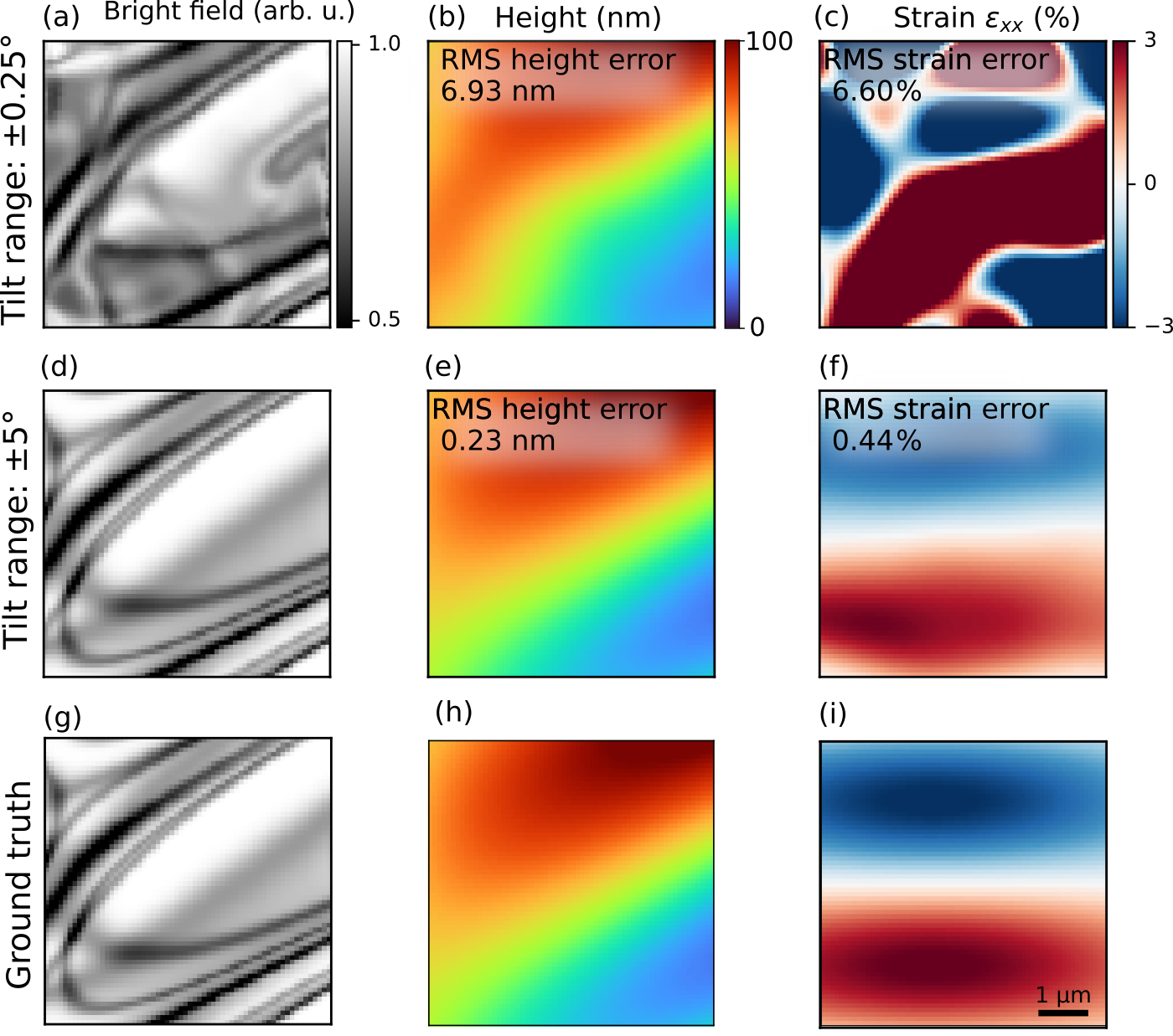}
    \caption{\textbf{Representative reconstructions at the extremes of tilt range.} (a--c)~Reconstructed bright-field image, height map, and $\varepsilon_{xx}$  strain map for a tilt range of $\pm 0.25^\circ$, in the non-unique reconstruction regime. (d--f)~Corresponding results for the full $\pm 5.0^\circ$ tilt range used in the experimental reconstruction. (g--i)~Intensity, height, and $\varepsilon_{xx}$ strain map sharing the same morphology as the surface in Fig.~\ref{fig:reconstruct_accuracy}. At $\pm 0.25^\circ$, the bright-field image appears qualitatively reasonable, but the height RMS error is as large as 6.93~nm and the strain RMS error is 6.60\%, with the strain map showing unphysical sharp features and saturation at the colormap limits. At $\pm 5.0^\circ$, both the height (RMS error 0.23~nm) and strain (RMS error 0.44\%) are accurately recovered, with smooth, physically meaningful spatial variation. The contrast between these two cases illustrates how an insufficient tilt range produces a loss landscape with multiple spurious minima that the optimizer can converge to, yielding low loss but a high ground-truth error.}
    \label{fig:supp:BF_comparison_tilt_range}
\end{figure}

 The qualitative difference between the non-unique and unique regimes is made explicit in Fig.~\ref{fig:supp:BF_comparison_tilt_range}, which compares representative reconstructions at $\pm 0.25^\circ$ and $\pm 5^\circ$ tilt range. At $\pm 0.25^\circ$, the reconstructed bright-field image is visually plausible for a large part within the field of view [Fig.~\ref{fig:supp:BF_comparison_tilt_range}(a)], yet the recovered height and strain maps are inconsistent with the ground truth, with an RMS height error of $6.93$~nm and an RMS strain error of $6.60\%$ [Fig.~\ref{fig:supp:BF_comparison_tilt_range}(b,c)]. The strain map of $\varepsilon_{xx}$, in particular, shows unphysical, sharply varying features. At $\pm 5^\circ$, the same reconstruction procedure recovers smooth, physically meaningful height and strain fields with RMS errors of $0.23$~nm and $0.44\%$ [Fig.~\ref{fig:supp:BF_comparison_tilt_range}(e,f)]. 

\begin{figure*}[htb!]
    \centering
    \includegraphics[width=0.8\linewidth]{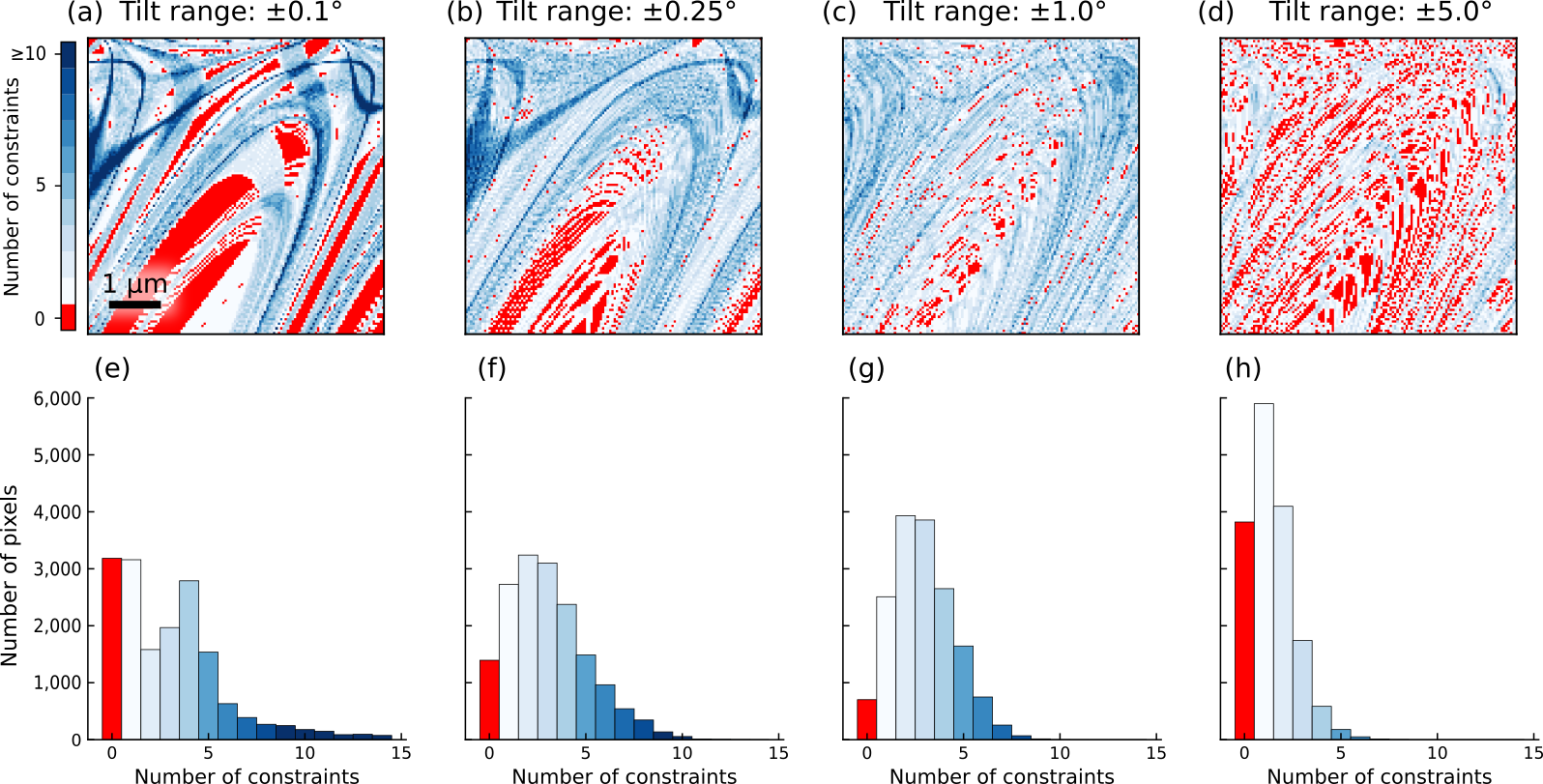}
    \caption{\textbf{Spatial coverage of bend contour constraints as a function of the tilt range.} (a--d)~Spatially-resolved maps of the number of bend contours that pass through each pixel for different tilt ranges, accumulated over a tilt series of 20 images. The color scale indicates the number of constraints per pixel, with red marking pixels traversed by no bend contour (zero constraints) and progressively darker blue indicating more constraints. As the tilt range increases initially, more of the field of view is covered by at least one contour, and pixels accumulate constraints from multiple nonparallel orders. (e--h)~Histograms of the number of constraints per pixel corresponding to the maps in (a--d), respectively. With a small tilt range in (a,~e), many pixels receive zero constraints (red bar) and few receive more than one, leaving the surface largely underdetermined. By a $\pm 1.0^\circ$ tilt range in (c,~g), the zero-constraint population is substantially reduced and the distribution shifts toward multiple constraints per pixel, providing the redundant, multi-angle coverage required for a well-posed reconstruction. At a $\pm 5.0^\circ$ tilt range in (d,~h), the zero-constraint population grows again because the tilt step is too large so that bend contours move large distances between successive tilts. Some small regions are still unconstrained in (f,~g), but by utilizing spatial regularization, the correct height/strain can still be reconstructed in these regions using neighboring constraints.}
    \label{fig:tilt_range_constraints}
\end{figure*}

As with the image-count analysis, what is important for identifiability of the inverse problem is the number of constraints gained from the tilt series. In Fig.~\ref{fig:tilt_range_constraints}(a--d), we plot the number of bend contours that traverse each pixel. At the $\pm0.1^\circ$ tilt range, many large regions of the surface do not have any bend contours, noted by the red pixels. This lack of constraints leads to an unreliable reconstruction. For higher tilt ranges, more bend contours traverse the field of view, and the constraints are more evenly distributed, leading to a constrained surface. It is clear from Fig.~\ref{fig:tilt_range_constraints}(a--d) that a tilt range of $\pm1^\circ$ is approximately ideal for this example surface because the bend contour constraints cover almost the entire surface, with a large swathe of the surface constrained by more than one contour [Fig.~\ref{fig:tilt_range_constraints}(g)]. By a tilt range of $\pm5^\circ$, the surface is relatively less constrained because the tilt step is too large; bend contours move large distances between tilts, again leading to some unconstrained regions. In an experiment, one can choose the minimal tilt range and number of images by making sure a few non-parallel bend contours traverse the entire field of view, and that the bend contours do not jump large distances between nearby tilt images in order to get a well-constrained solution.

\subsection{Minimum image number required for unique reconstruction and implications on ultrafast TEM}

Whether a BCET reconstruction is well-constrained is governed \textit{jointly} by the tilt range and the number of images. While choosing the tilt range to minimize the image count and to still find a quantitative reconstruction, we find that a well-constrained solution can be obtained with as few as 5~tilt images for the example surface in Fig.~\ref{fig:supp:BF_comparison_tilt_range}, provided that a tilt range of $\pm0.5^\circ$ is used. Under these conditions, reconstructions from randomized starting points remain tightly distributed with low RMS errors of 0.5\% and 2~nm for strain and height, respectively.

For ultrafast TEM measurements where data acquisition takes much longer than static TEM experiments, the minimum data requirement can be further relaxed by a few regularization strategies. First, a complete tilt series can be acquired at the pre-time-zero pump-probe delay time to establish an accurate reference reconstruction of the pre-excitation surface height and strain fields. For a typical micrometer-sized film, due to the constraint of typical sound speeds, its morphology cannot change substantially between adjacent pump--probe time delays on the pico- to nanosecond timescale. Hence, this pre-excitation reference serves as a strong prior that constrains the space of physically admissible solutions at each time frame after time-zero. Second, the structural response of the film to an ultrafast pump can itself be physically constrained. For instance, in the absence of phase transitions, small-amplitude, elastic thin film deformations following laser excitation take the form of well-defined Lamb wave modes or specific acoustic wave packets, whose displacement and strain fields conform to known analytical forms \cite{du_Imaging_2020}. These physical priors can be incorporated directly into the iterative reconstruction at each time delay. For example, we can restrict the allowed perturbations in time to a low-dimensional basis of physically motivated displacement fields. This strategy dramatically reduces the effective number of free parameters and the minimum number of tilt images required for converging to a unique solution. Together, these two factors make BCET well suited to the data-limited conditions of ultrafast pump--probe measurements, where only one or a handful of tilt images may be available per time delay.

\begin{figure*}[htb!]
    \centering
    \includegraphics[width=0.75\linewidth]{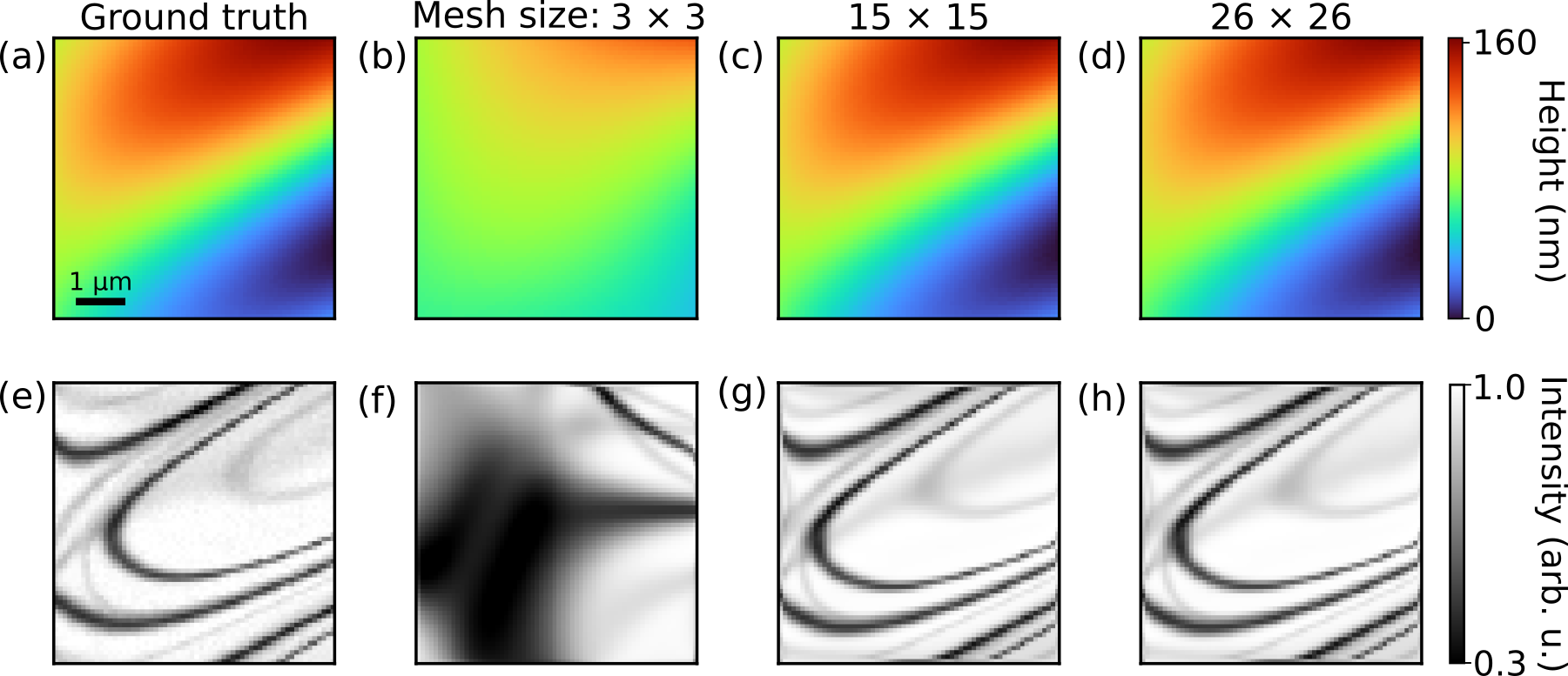}
    \caption{\textbf{Reconstruction of a simulated monoclinic Ga$_\text{2}$O$_\text{3}$ film by the direct solver.} (a)~Ground truth height map. (b--d)~Reconstructed height map for the direct solver with different control point mesh sizes. At around the mesh size of 15 by 15 control points, the surface morphology is well reconstructed. (e--h)~Bright field images at the same sample tilt angle of $\theta_\alpha=\theta_\beta=0^\circ$ for each of the height maps in (a--d), respectively.}
    \label{fig:direct_monoclinic}
\end{figure*}

\section{R\lowercase{econstruction of a low-symmetry crystal}}
\label{supp:low_symmetry_recon}

\begin{figure*}[htb!]
    \centering
    \includegraphics[width=0.8\linewidth]{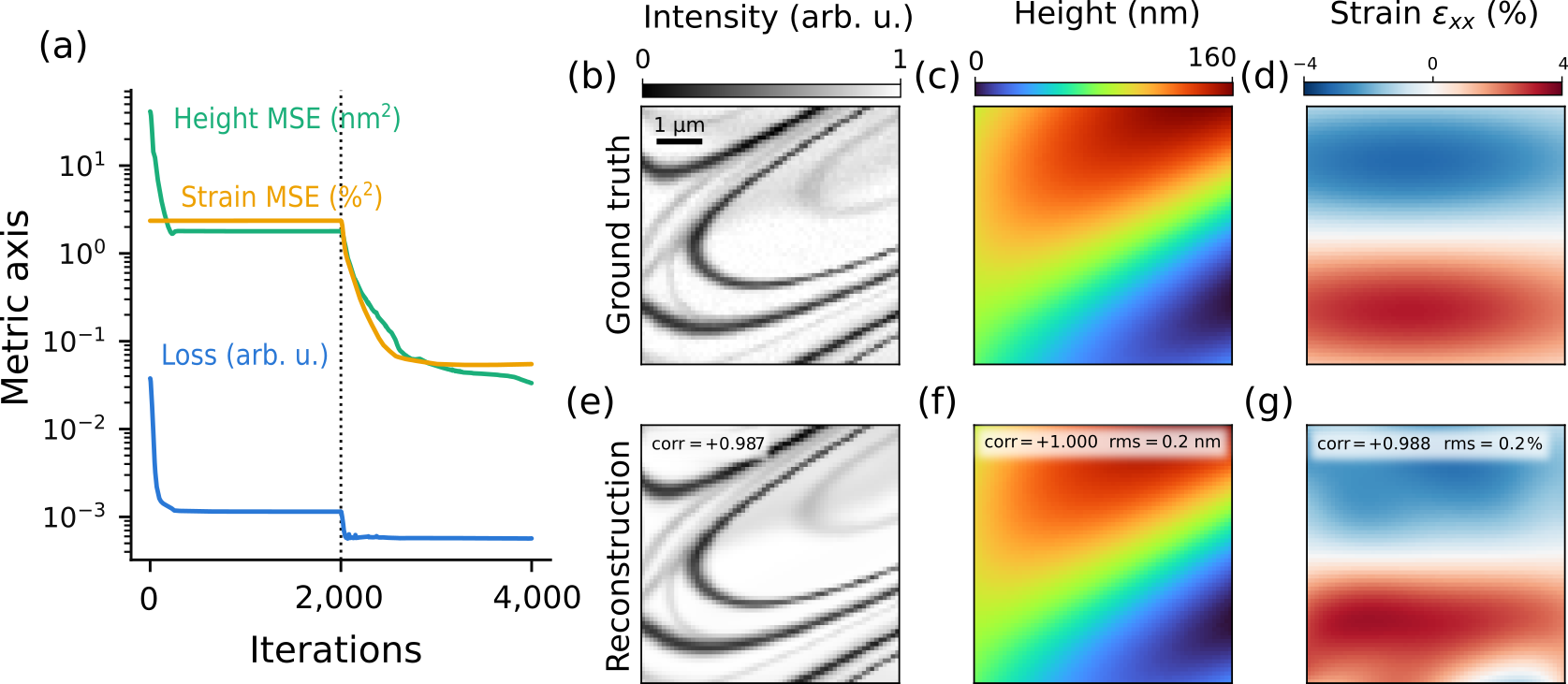}
    \caption{\textbf{Reconstruction of a simulated monoclinic Ga$_\text{2}$O$_\text{3}$ film by our iterative procedure.} (a)~Loss, height mean squared error (MSE), and strain MSE for the iterative reconstruction. The initial guess for the reconstruction is a randomly sampled surface. We used a two-stage reconstruction in which the strain degrees of freedom were held fixed for the first 2,000 iterations and released thereafter, marked by the dashed vertical line. (b)~Bright field image with added Poisson noise matching experimental conditions (8,200 electrons/pixel). (c,~d)~Height and $\varepsilon_{xx}$ strain maps for the ground truth, respectively. (e)~Reconstructed surface bright field image with a normalized cross-correlation score of 0.987 with the ground truth image. (f)~Reconstructed height map, which has a correlation of 1.000 and an RMS error of 0.2~nm. (g)~Reconstructed strain map with a correlation of 0.988 and an RMS error of 0.2\%.}
    \label{fig:iterative_monoclinic}
\end{figure*}

The formalism underlying both the direct and iterative solvers applies to an arbitrary unit cell and is not restricted to cubic or otherwise high-symmetry crystal systems exemplified by SrTiO$_3$ presented in the main text. As derived in Sec.~\ref{supp:general_direct_constraints}, the general direct-solver constraint [Eq.~\eqref{eq:direct_method_constraint}] is written in terms of the angle $\psi_{hkl}$ between the reciprocal lattice vector $\mathbf{G}_{hkl}$ and the local surface normal $\hat{\mathbf{n}}$, which is computed directly from the lattice parameters $a,b,c$ and inter-axial angles $\alpha,\beta,\gamma$ of a general unit cell through Eqs.~\eqref{eq:normal_vector}--\eqref{eq:direct_method_constraint}. The cubic constraint used for the experimental SrTiO$_3$ reconstruction in the main text [Eq.~\eqref{eq:cubic_constraint}] is simply the special case $\psi_{hkl} = \pi/2$, which follows when $\alpha = \beta = \gamma = \pi/2$ and $l = 0$. Likewise, the iterative forward model computes the reciprocal lattice vectors $\mathbf{G}_{hkl}(u,v)$ directly from the deformed local coordinate matrix $S(u,v)$ (see Sec.~\ref{supp:surf_param}), so lowering the crystal symmetry changes only the numerical values of the lattice matrix $M$ and introduces no additional terms or computational cost in the optimization procedure.

To demonstrate this generality of BCET applied to low-symmetry crystalline films, we performed simulated reconstructions using a monoclinic crystal, $\beta$-Ga$_2$O$_3$ (space group $C2/m$, No.~12), with lattice parameters $a = 12.21$~\AA, $b = 3.04$~\AA, $c = 5.80$~\AA{}, $\alpha = \gamma = 90^\circ$, and $\beta = 103.8^\circ$ \cite{ahman_Reinvestigation_1996}. We used the same ground-truth topography and strain field, initial guess, and optimization procedure as in the cubic validation of Sec.~\ref{supp:accuracy}, changing only the crystal system used to generate the bend contours. These measures isolate the effect of crystal symmetry on reconstruction performance from all other factors.

We first applied the direct solver to recover the strain-free surface topography, using the general angular constraint [Eq.~\eqref{eq:direct_method_constraint}] appropriate for the monoclinic cell. Figure~\ref{fig:direct_monoclinic} shows the ground-truth height map alongside direct-solver reconstructions using B\'ezier control-point meshes of increasing sizes. As with the cubic case (see Fig.~\ref{fig:test_direct}), a coarse $3 \times 3$ mesh underfits and fails to capture the surface curvature [Fig.~\ref{fig:direct_monoclinic}(b,f)], while a $15 \times 15$ or $26 \times 26$ mesh accurately recovers both the height map and the corresponding bright-field bend-contour pattern [Fig.~\ref{fig:direct_monoclinic}(c,d,g,h)]. The direct solver therefore supplies a suitable initial estimate for the iterative reconstruction once a sufficient control-point basis is provided.

We then performed the full iterative reconstruction, starting from a randomly sampled initial surface and using bright-field images with Poisson noise matching our experimental conditions (8,200 electrons/pixel; see Sec.~\ref{supp:accuracy}). The loss, height error, and strain error all converge to small values over the course of the optimization [Fig.~\ref{fig:iterative_monoclinic}(a)]. The converged reconstruction recovers the ground-truth topography and in-plane strain tensor with accuracy comparable to the cubic case: the reconstructed height map has a correlation of $1.000$ and an RMS error of $0.2$~nm relative to the ground truth [Fig.~\ref{fig:iterative_monoclinic}(f)], while the reconstructed $\varepsilon_{xx}$ strain map has a correlation of $0.988$ and an RMS error of $0.2\%$ [Fig.~\ref{fig:iterative_monoclinic}(g)]. The reconstructed bright-field image reproduces the ground-truth bend-contour pattern with a correlation of $0.987$ [Fig.~\ref{fig:iterative_monoclinic}(e)].
 These results confirm that reduced crystal symmetry does not degrade the performance of either the direct or the iterative solver in BCET.

\begin{figure*}[htb!]
    \centering
    \includegraphics[width=0.64\linewidth]{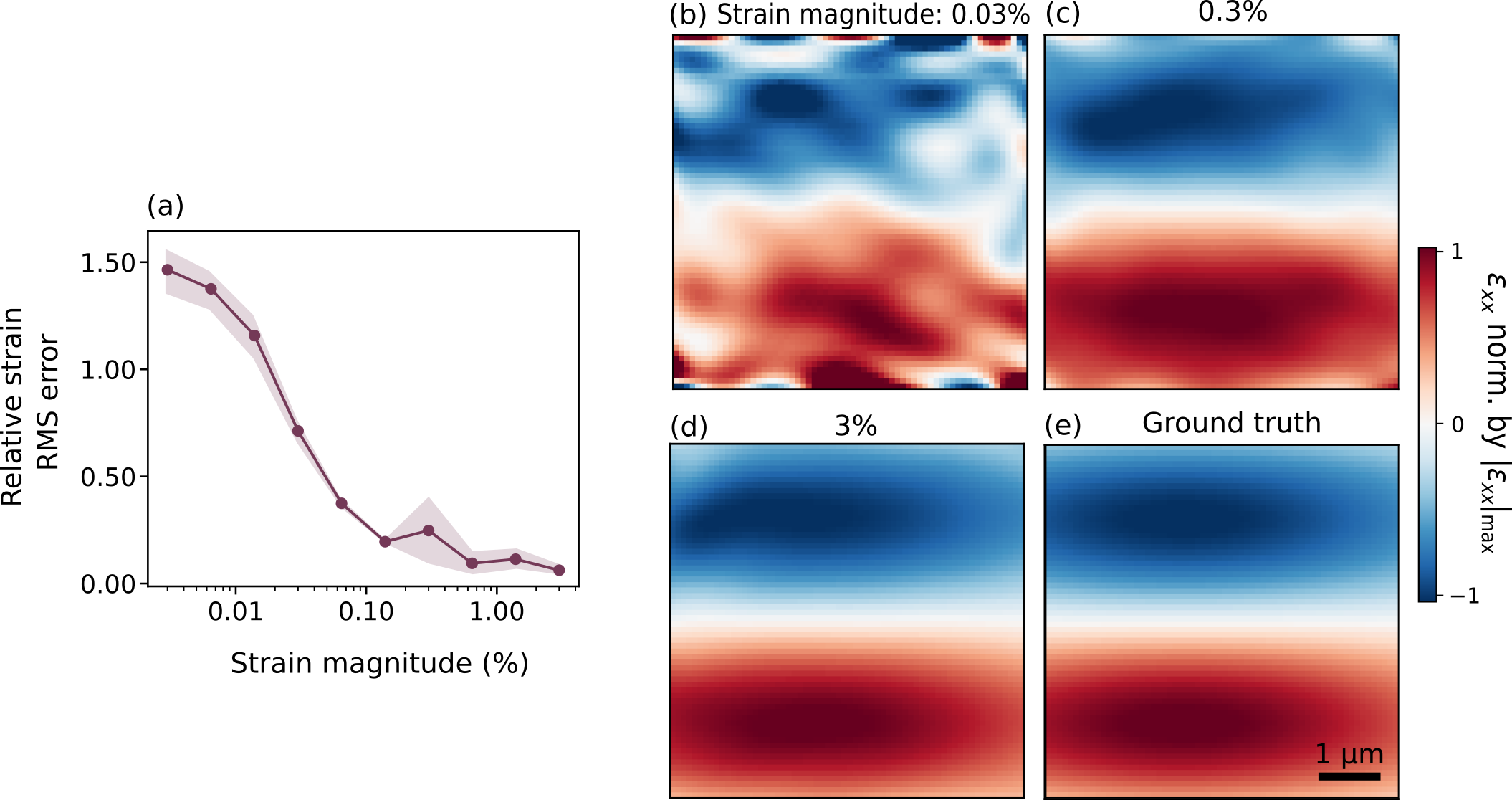}
    \caption{\textbf{Simulated iterative reconstruction as a function of ground-truth strain magnitude.} (a)~Average relative strain RMS error, defined as the RMS error between the reconstructed and ground-truth strain normalized by the RMS of the ground-truth strain. The average is taken over 25~reconstruction runs at each strain magnitude, and the shaded band represents the standard error across reconstructions. The relative error rises sharply below approximately 0.1\% strain, indicating the onset of noise-dominated reconstruction. (b--d)~Representative reconstructed $\varepsilon_{xx}$ strain maps for ground-truth strain magnitudes of 0.03\%, 0.3\%, and 3\%, each normalized by the maximum strain of the corresponding ground-truth surface. (e)~The ground truth strain map normalized by the maximum strain. At 3\% and 0.3\% strain values, the reconstructed strain field faithfully reproduces the ground-truth spatial pattern, whereas at 0.03\% strain, the reconstruction is dominated by noise and is on the border of recovering the underlying strain distribution.}
    \label{fig:strain_sensitivity}
\end{figure*}

\section{S\lowercase{train sensitivity analysis}}
\label{supp:strain_sensitivity}

Having established the accuracy and precision of BCET for a fixed strain field in Secs.~\ref{supp:precision} and~\ref{supp:accuracy}, we now examine how the strain reconstruction degrades as the magnitude of the underlying strain field is reduced. This analysis establishes an approximate lower bound on the strain that BCET can reliably recover under realistic noise conditions, and complements the noise scaling analysis of Sec.~\ref{supp:accuracy}, where the strain magnitude was held fixed and the electron count was varied.

To isolate the effect of strain magnitude, we used the same ground-truth topography, initial guess, and optimization procedure as in the accuracy validation of Sec.~\ref{supp:accuracy}, and scaled the in-plane strain field of the ground-truth surface to a series of target RMS magnitudes spanning $0.03\%$ to $3\%$. For each strain magnitude, the corresponding bright-field tilt series was simulated with Poisson noise matching our experimental conditions (8,200 electrons/pixel; see Sec.~\ref{supp:accuracy}), and we performed 25 independent reconstructions from randomly sampled starting points. Because the absolute strain RMS error is not directly comparable across surfaces with different strain magnitudes, we quantify performance using the \emph{relative} strain RMS error, defined as the RMS error between the reconstructed and ground-truth strain normalized by the RMS of the ground-truth strain. A relative error near zero indicates faithful recovery, while a relative error of order unity indicates that the reconstructed strain bears little resemblance to the ground truth.

The average relative strain RMS error over the 25 reconstructions is shown as a function of strain magnitude in Fig.~\ref{fig:strain_sensitivity}(a). For strain magnitudes of $\sim\!0.1$\% and above, the relative strain error remains small ($< 0.25$) and the reconstructed strain maps faithfully reproduce the ground-truth spatial pattern [Fig.~\ref{fig:strain_sensitivity}(c,d)]. As the strain magnitude is reduced below 0.1\%, the relative error rises steeply and the spread across reconstructions widens, reflecting the transition to a noise-dominated regime in which the strain-induced modulation of the bend-contour positions becomes comparable to the noise floor. As an example in this regime, at 0.03\% strain, the reconstruction struggles to recover the underlying strain distribution [Fig.~\ref{fig:strain_sensitivity}(b)].

Taken together, these results indicate that BCET reliably reconstructs in-plane strain fields down to a magnitude of approximately $0.1\%$ under the noise conditions of our experiment, with performance degrading below this scale. We emphasize that this threshold is not fundamental but depends on the experimental electron count, the tilt series, and the surface morphology; higher electron doses or larger tilt ranges would be expected to extend the sensitivity to smaller strains. The $\sim\!0.1\%$ sensitivity established here is well below the strain magnitudes of order $1\%$ recovered in our experimental SrTiO$_3$ reconstruction [Fig.~\ref{fig:3}(h--j)], confirming that the measured strain field lies comfortably within the reliably reconstructible regime.
\color{black}

\begin{figure}[t!]
    \centering
    \includegraphics[width=0.75\linewidth]{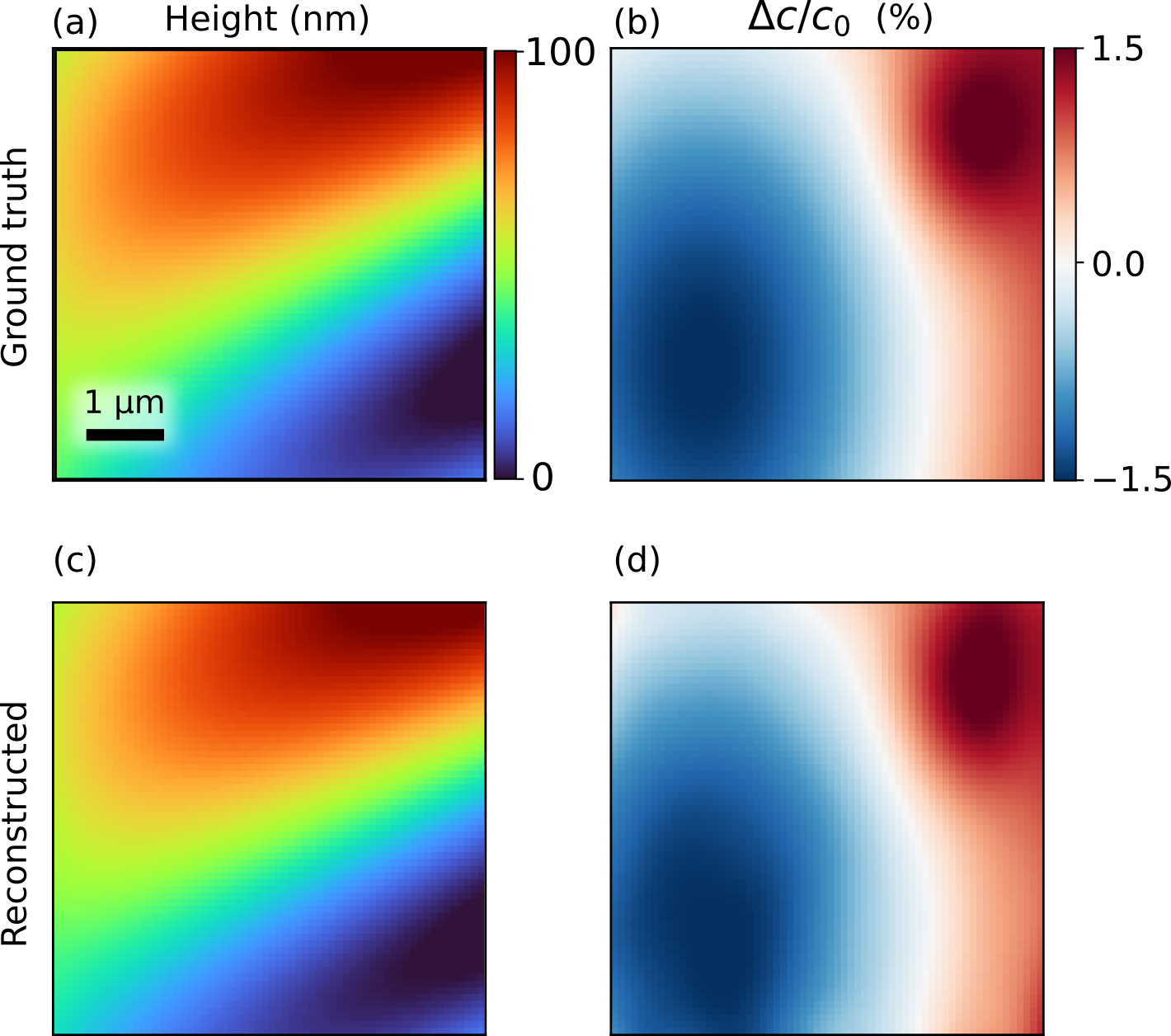}
   \caption{\textbf{Simulated reconstruction of the out-of-plane lattice strain.} (a,~b)~Spatial maps of the ground truth height and the variation of the $c$ lattice parameter; $c_0$ denotes the unstrained value. (c,~d)~Reconstructed height and $c$ lattice parameter maps. The reconstruction uses 12 images with varying $\theta_\alpha \in [14^\circ, 24^\circ]$ while fixing $\theta_\beta = 0^\circ$, and another 12 images fixing $\theta_\alpha = 19^\circ$ while varying $\theta_\beta \in [-5^\circ, 5^\circ]$. The ground-truth height is the same as that used in Fig.~\ref{fig:reconstruct_accuracy}.}
    \label{fig:c_lattice_recon}
\end{figure}

\section{O\lowercase{ut-of-plane strain reconstruction}}
\label{supp:c_lattice_recon}

The implementation of BCET presented in the main text reconstructs the in-plane strain tensor and out-of-plane topography but assumes that the surface normal is a unit vector, hence neglecting the out-of-plane strain component. This restriction can be relaxed by allowing the surface normal to deviate from unit length, which would extend the formalism to capture spatially-dependent out-of-plane lattice variations. However, reconstructing the out-of-plane strain requires observation of bend contours beyond the zeroth-order Laue zone. For our current experimental dataset of SrTiO$_3$, we only observed $l = 0$ bend contours due to the limited tilt range. Depending on the tilt range of the holder, it is possible to reach contours from higher-order Laue zones. 

To explicitly demonstrate this capability, we have performed a simulated reconstruction using a tilt range with $\theta_\alpha$ elevated to $19^\circ$ to reach a higher-order Laue zone. Using 24~tilt images with a range of $\pm 5^\circ$ in both tilt axes, the ground-truth map of the $c$ lattice parameter was successfully reconstructed (see Fig.~\ref{fig:c_lattice_recon}).

\color{black}

\section{S\MakeLowercase{ample preparation} \label{supp:sample_prep}} 
Single-crystal SrTiO$_3$ (STO) films were epitaxially grown on the (001) surface of SrTiO$_3$ substrates by pulsed laser deposition (PLD) using a KrF excimer laser ($\lambda = 248$~nm, Coherent) and a polycrystalline ceramic STO target. The base pressure of the PLD chamber was $5 \times 10^{-9}$~Torr. Prior to deposition, the STO(001) substrates were annealed at $1000^{\circ}\mathrm{C}$ for 1~hour in $1 \times 10^{-6}$~Torr of oxygen to obtain a clean and atomically flat surface. A 10-nm-thick Sr$_2$CaAl$_2$O$_6$ (SCAO) sacrificial layer was then deposited at $700^{\circ}\mathrm{C}$ under an oxygen pressure of $5 \times 10^{-6}$~Torr, with its thickness calibrated by reflection high-energy electron diffraction oscillations. During the growth, the laser repetition rate was 1~Hz and the fluence was 1.14~J/cm$^2$. Subsequently, a 30-nm-thick STO film was deposited on the SCAO layer at $700^{\circ}\mathrm{C}$ under an oxygen pressure of $5 \times 10^{-6}$~Torr, using a laser repetition rate of 2~Hz and a fluence of 0.85~J/cm$^2$. After growth, the SCAO layer was dissolved in deionized water to release the STO film, which was subsequently transferred onto a 2,000-mesh copper grid for TEM measurements.

\bibliography{references}

\end{document}